\documentclass[fleqn,usenatbib,useAMS]{mnras}
\usepackage[dvipsnames]{xcolor}
\usepackage{tikz,hyperref}

\definecolor{lime}{HTML}{A6CE39}
\DeclareRobustCommand{\orcidicon}{
	\begin{tikzpicture}
	\draw[lime, fill=lime] (0,0) 
	circle [radius=0.13] 
	node[white] {{\fontfamily{qag}\selectfont \tiny ID}};
	\draw[white, fill=white] (-0.0625,0.095) 
	circle [radius=0.007];
	\end{tikzpicture}
	\hspace{-2mm}
}

\foreach \x in {A, ..., Z}{\expandafter\xdef\csname orcid\x\endcsname{\noexpand\href{https://orcid.org/\csname orcidauthor\x\endcsname}
			{\noexpand\orcidicon}}
}
\usepackage{newtxtext,newtxmath}

\usepackage{booktabs}
\usepackage[T1]{fontenc}
\usepackage{ae,aecompl}

\usepackage{graphicx}	% Including figure files
\usepackage{amsmath}	% Advanced maths commands
\usepackage{booktabs}
\usepackage{times}
\input{epsf}
\title[Variable X-ray emission from T CrB]{The X-ray rise and fall of the Symbiotic Recurrent Nova System T~CrB}

\author[J.~A.~Toal\'{a} et al.]{Jes\'{u}s~A.~Toal\'{a}\thanks{E-mail:\,j.toala@irya.unam.mx}\thanks{Visiting astronomer at the IAA-CSIC as part of the Centro de Excelencia
Severo Ochoa Visiting-Incoming programme.}$^{1}\orcidA$, Omaira~Gonz\'{a}lez-Mart\'{i}n$^{1}\orcidB$, Andrea Sacchi$^{2}\orcidC$ and Diego~A.~Vasquez-Torres$^{1}\orcidD$
\\
$^{1}$Instituto de Radioastronom\'{i}a y Astrof\'{i}sica, Universidad Nacional Aut\'{o}noma de M\'{e}xico, 58089 Morelia, Michoac\'{a}n, Mexico\\
$^{2}$Center for Astrophysics $\vert$\ Harvard \& Smithsonian, 60 Garden Street, Cambridge, MA 02138, USA
}

\date{\today}%Accepted XXX. Received YYY; in original form ZZZ}

\pubyear{2024}

\begin{document}
\label{firstpage}
\pagerange{\pageref{firstpage}--\pageref{lastpage}}
\maketitle

\begin{abstract}
We present the analysis of publicly available {\it NuSTAR}, {\it Suzaku} and {\it XMM-Newton} observations of the symbiotic recurrent nova T~CrB covering the 2006.77--2022.66 yr period. The X-ray spectra are analysed by adopting a model that includes a reflection component produced by the presence of a disk that mimics the accretion disk and the immediate surrounding medium. Our best-fit model requires this disk to have a radius of 1 AU, effective thickness of 0.1 AU, averaged column density 10$^{25}$~cm$^{-2}$ and orientation of 50$^{\circ}$ with respect to the line of sight. This disk is about a factor of two larger than recent estimations for the accretion disk and its presence contributes significantly  via reflection to the total X-ray flux detected from T~CrB, which  naturally produces the emission of the 6.4 keV Fe line. Our analysis suggests that the temperature of the boundary layer evolved from 14.8 keV in the steady-state phase (before 2016), to 2.8 keV in the 2017.24 epoch, to finally stabilise to about $\sim$8 keV in the subsequent epochs.  These variations in the plasma temperature of the boundary layer are attributed to the evolution of the mass accretion rate ($\dot{M}_\mathrm{acc}$), which is estimated to have an averaged value of $\dot{M}_\mathrm{acc}$=2.6$\times10^{-8}$~M$_\odot$~yr$^{-1}$ for the current active phase. The presence of emission lines in the {\it XMM-Newton} RGS spectrum of 2017.24 prevents from adopting a black body emission model to fit the soft X-ray range. Instead, we use plasma emission models that suggest the presence of adiabatically-shocked gas produced by gas velocities of 110--200~km~s$^{-1}$, very likely tracing jet-like ejections similar to what is found in other symbiotic systems. The analysis of X-ray and optical data together show that T~CrB has a similar evolution as black hole binaries, accreting neutron stars and AGN in the hardness-intensity diagram.
\end{abstract}


%%%%%%%%%%%%%%%%%%%%%%%%%%%%%%%%%%%%%%%%%%%%%%%%%%

%%%%%%%%%%%%%%%%% BODY OF PAPER %%%%%%%%%%%%%%%%%%

\section{Introduction}\label{introduction}
\label{sec:intro}

There is a variety of accreting white dwarf (WD) systems producing different and interesting physical phenomena \citep[see][]{Mukai2017,Webb2023}. They are the progenitors of supernova Type Ia and, consequently, make them key astronomical objects not only for stellar astrophysics but also for cosmological purposes. Among these, symbiotic stars comprise an accreting WD and red giant companion capable of producing any kind of electromagnetic signature \citep[see][]{Luna2013}. Regardless of the accretion process \citep[Bondi-Hoyle accretion, Roche-Lobe overflow, or a hybrid wind Roche-lobe overflow channel; e.g.,][]{Bondi1944,Podsiadlowski2007} it has been suggested that at least 80 per cent of them harbour an accretion disk \citep{Merc2024}, which might also power the launching of non-relativistic jets through the so-called jet-feedback mechanism \citep[e.g.,][]{Soker2016}.

X-ray emission has been used as an invaluable tool to study the accretion physics in symbiotic stars. 
The variety of X-ray spectra detected from symbiotic systems has been explained by a combination of different factors. In a pioneer work, \citet{Murset1997} used {\it ROSAT} observations of symbiotic stars to propose the first classification scheme using X-ray data. $\alpha$-type symbiotic stars were defined as super soft ($E<0.5$~keV) sources, typically attributed to the thermonuclear burning on the surface of the WD \citep[e.g.,][]{Orio2007}. $\beta$-type objects have spectra that peak at 0.8~keV produced by plasma with temperatures of $\sim$10$^{6}$~K, attributed to colliding winds, accretion shocks, and/or
accretion disk. Finally, \citet{Murset1997} defined $\gamma$-type objects exhibiting harder X-ray emission with emission up to 2.4~keV (the {\it ROSAT} energy limit), that are now associated with symbiotic systems where the compact object is a neutron star and which spectra are modelled with power law components \citep[non-thermal origin;][]{Merc2019}.

With the advent of subsequent generation of X-ray instruments, it was demonstrated that symbiotic stars also emit harder ($E>10$~keV) X-rays \citep[e.g.,][]{Chernyakova2005,Kennea2009} and the $\alpha/\beta/\gamma$ classification scheme needed to be revised.
\citet{Luna2013} proposed a $\delta$-type which corresponded to highly-absorbed, hard X-ray-emitting symbiotic stars. This hard X-ray emission is very likely produced at the innermost accretion region around the WD, at the {\it boundary layer} between the WD's surface and the inner region of the accretion disk. Moreover, some other sources exhibit more complex spectra, characterised by the presence of both soft and hard X-ray emission and, thus, a mixed classification $\beta/\delta$-type is used \citep[see][and references therein]{Merc2019,Luna2013}.

\begin{table*}
\caption{Details of the publicly available {\it XMM-Newton} observations of T~CrB used here. All observations were obtained with the full-frame mode and the medium optical blocking filter. The count rates were obtained for the complete (0.3--10.0~keV) and soft (0.3--2.0~keV) energy ranges.}
\setlength{\tabcolsep}{0.8\tabcolsep}  
\label{tab:obs}
\begin{center}
\begin{tabular}{ccccccccccccc}
\hline
Revolution & Obs.~ID. & Observation start & \multicolumn{3}{c}{Total exposure time} & \multicolumn{1}{c}{Useful Time} & \multicolumn{1}{c}{count rate} & soft count rate \\
\cmidrule(lr){4-6} \cmidrule(lr){7-7} \cmidrule(lr){8-8} \cmidrule(lr){9-9}
       &             &                   &  pn   & MOS1   & MOS2 & pn   &  pn             & pn  \\
       &             &  (UTC)            & (ks)  & (ks)   & (ks) & (ks) & (cnt~ks$^{-1}$) & (cnt~ks$^{-1}$)\\
\hline
3152 & 0793183601 & 2017-02-23T03:47:34 & 60.8  & 62.4 & 62.5 & 42.4 &  117$\pm$2 & 94$\pm$2 \\
3323 & 0800420201 & 2018-01-30T04:25:43 & 25.0  & 26.6 & 26.7 & 20.9 &  112$\pm$3 & 36$\pm$2 \\ 
3800 & 0864030101 & 2020-09-07T23:50:02 & 50.6  & 52.4 & 52.4 & 32.5 &  94$\pm$2  & 5$\pm$1  \\
3975 & 0882640301 & 2021-08-23T03:07:24 & 57.9  & 62.6 & 62.6 & 44.9 &  248$\pm$3 & 5$\pm$1  \\
4074 & 0882640401 & 2022-03-08T09:52:59 & 49.8  & 51.6 & 51.6 & 30.8 &  184$\pm$3 & 32$\pm$1 \\
4146 & 0882640601 & 2022-07-29T17:47:04 & 53.6  & 51.8 & 55.5 & 36.3 &  407$\pm$4 & 11$\pm$1 \\
\hline
\end{tabular}
\end{center}
\end{table*}

Multi-epoch X-ray observations have shown that some symbiotic stars exhibit dramatic changes in their spectra \citep[e.g.,][]{Pujol2023,Yu2022}. For example, \citet{Toala2023} showed that the 2016 {\it XMM-Newton} spectrum of HM~Sge presented a soft ($<$0.5~keV) component previously undetected in {\it ROSAT} data \citep{Murset1997}. That is, the previous $\beta$-type classification assigned to HM~Sge no longer holds. Furthermore, the extreme variability of the jets can produce detectable variability at energies below $E<0.5$ keV \citep[see][]{Sacchi2024}.

X-ray variability has also been reported in one of the most studied symbiotic stars, the $\beta/\delta$-type CH~Cyg. Multi-epoch, multi-instrument observations showed that its total luminosity has changed by an order of magnitude in the past decades \citep[see, e.g.,][and references therein]{Mukai2007}.
In \citet{Toala2023b} our group presented one of the most complete studies of this $\beta/\delta$-type symbiotic stars. High-resolution X-ray spectra were used to accurately determine the abundances of the X-ray-emitting gas as well as to confirm the presence of multi-temperature gas using the emission from He-like triplets. In that work, we demonstrated that the presence of a reflection component produced by an ionised disk is important to model the medium energy range (2.0--4.0~keV) and to appropriately fit the presence of the fluorescent Fe emission line at 6.4~keV (without the need of adding a Gaussian component). In addition, in the best model of CH~Cyg the reflection component also contributes to $\sim$10\% of the soft X-ray emission. A comparison with previous works suggested that the variation of the Fe K$\alpha$ emission line at 6.4 keV is correlated with possible variations in the reflection component.

As a consequence of these works, we have started a series of papers to address the X-ray properties of the reflection component in highly variable symbiotic stars. We want to test further our ionised disk model to try to push forward our understanding of the accretion and reflection processes in symbiotic systems. Here we present the analysis of publicly available X-ray observations of T~CrB. This is a recurrent symbiotic star in which the secondary is a M4III star \citep{Murset1999}. It has a period is 227.6~d \citep[][]{Kenyon1986,Fekel2000} and its distance has been estimated to be $d$=896$\pm$22~pc using {\it Gaia} data \citep{BailerJones2021}\footnote{Parallax distance measurements of symbiotic stars are affected by significant uncertainties given their extension, colours and orbital motion \citep[see, e.g.,][]{Linford2019,Sion2019}, however, with all this caveats, the one reported is the best available estimate of T~CrB distance. Other distance estimates range up to 1 kpc \citep{Hric1998}, which might affect less than 25 percent the luminosity estimations.}

T~CrB was first detected in X-rays by the {\it Einstein} observatory \citep{Cordova1981} and has been routinely observed by more modern X-ray instruments. The first hard X-ray detection of T~CrB during its quiescent state was reported by \citet{Tueller2005}, but subsequent analysis of the {\it Swift} and {\it Suzaku} data suggested the presence of heavily-absorbed, high-temperature plasma, producing hard X-ray emission \citep{Luna2008,Kennea2009}. This source was then classified as a $\delta$-type symbiotic star, but by early 2014 it started brightening in the optical, entering a "super active" phase \citep{Munari2016}. In fact, the optical light curve exhibited a peak around 2016 April 4--5 (2016.35) and subsequent {\it XMM-Newton} observations obtained during 2017 showed that its spectrum developed a bright soft component, which changed T~CrB classification into a $\beta/\delta$-type \citep{Luna2018}. Further analysis of 2017 and 2018 {\it XMM-Newton} observations in comparison with {\it Suzaku} and {\it Swift} data presented by \citep{Zhekov2019} showed that the flux of the soft component was declining while the opposite occurred to the hard component. \citet{Zhekov2019} estimated mass accretion rates ($\dot{M}_\mathrm{acc}$) for the different epochs and found that this was higher for the quiescent state than for the active phase (see fig. 6 of that publication).

T~CrB offers us a unique possibility to study the extreme changes of accretion states experienced by a symbiotic recurrent nova. In this paper, we present the analysis of multi-epoch (2006.77--2022.66), multi-instrument ({\it XMM-Newton, Suzaku and NuSTAR}) X-ray observations that cover the evolution from the steady-state phase up to the current active phase that, in combination with optical monitoring, can help us assess the role of the accretion process and the reflection physics. The analysis presented here is paving the road for the understanding of such an iconic recurrent symbiotic system on its way to producing a nova-like outburst in the near future.

This paper is organised as follows. In Section~\ref{sec:obs} we describe the observations and their processing methodology. Section~\ref{sec:results} presents the details of the spectral analysis and the results which are later discussed in Section~\ref{sec:discussion}. Finally, we present our conclusions in Section~\ref{sec:summary}.

\begin{figure*}
\begin{center}
\includegraphics[angle=0,width=\linewidth]{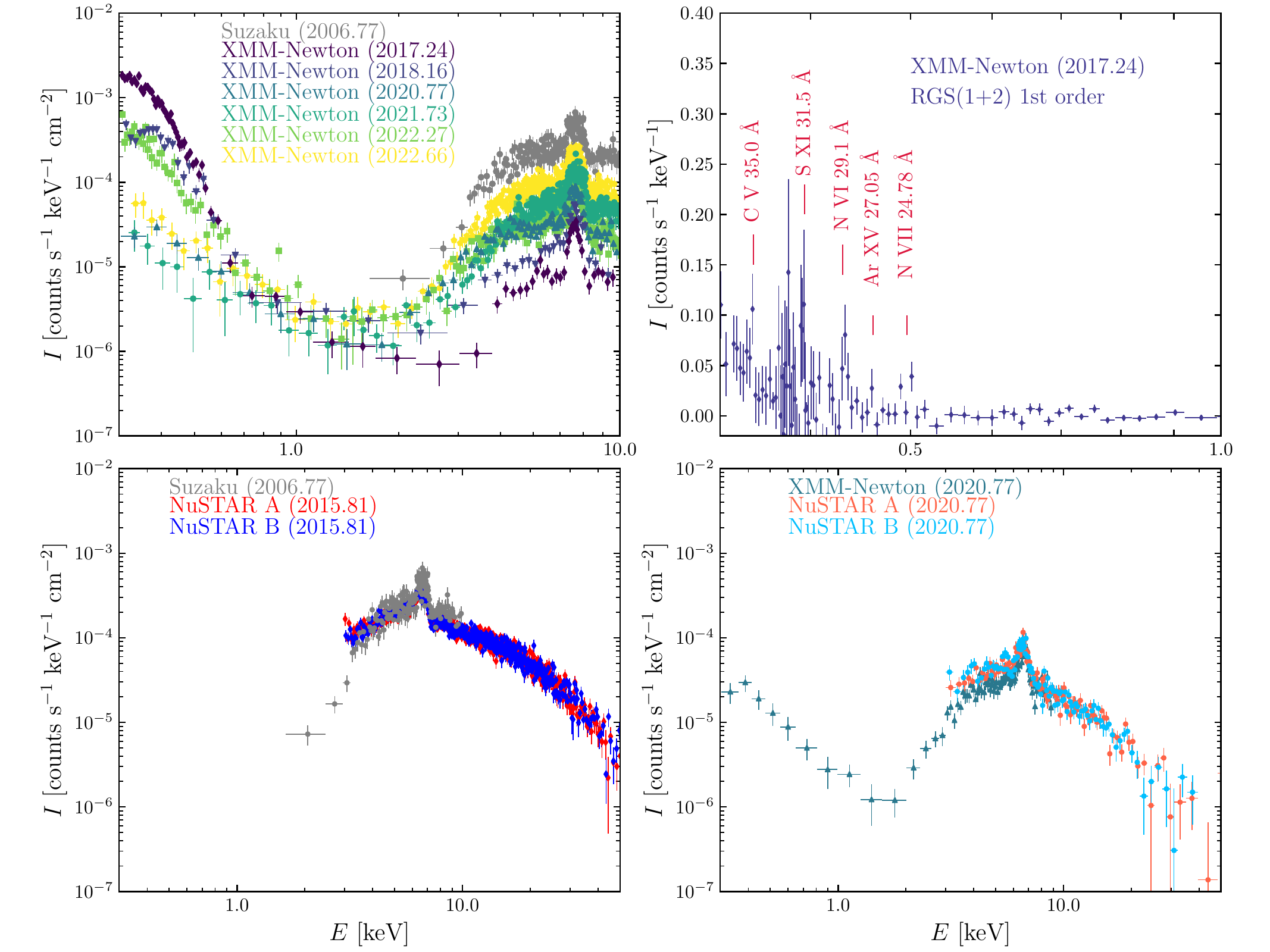}
\caption{Background-subtracted X-ray spectra of T~CrB. {\bf Top left:} Medium-resolution {\it XMM-Newton} EPIC-pn and {\it Suzaku} spectra. {\bf Top right:} high-resolution {\it XMM-Newton} RGS~(1+2) 1st order spectrum obtained from the 2017.24 {\it XMM-Newton} observations. The most prominent emission lines are labelled. {\bf Bottom:} {\it NuSTAR} spectra compared with {\it Suzaku} and {\it XMM-Newton} data. In all panels, the epochs are shown in years.}
\label{fig:spec1}
\end{center}
\end{figure*}

\section{Observations and data preparation}
\label{sec:obs}

\subsection{{\it XMM-Newton} observations}

T~CrB has been observed on several occasions by {\it XMM-Newton} using the European Photon Imaging Camera (EPIC), the Reflection Grating Spectrometers (RGSs) and the optical monitor (OM). T~CrB has been observed on 6 occasions, that correspond to Obs.~ID. 0793183601, 0800420201, 0864030101, 0882640301, 0882640401 and 0882640601. All these observation data files were retrieved from the XMM-Newton Science Archive\footnote{\url{http://nxsa.esac.esa.int/nxsa-web/\#search}}. We note that all of the EPIC observations were obtained using the full frame mode with the medium optical blocking filter. Details of the {\it XMM-Newton} observations are presented in Table~\ref{tab:obs}.

The {\it XMM-Newton} data were processed with the Science Analysis Software \citep[{\sc sas};][]{Gabriel2004} version 20.0 with the calibration files obtained on 2023 March 20.
The process follows standard {\sc sas} routines which include generating the EPIC even files with the \texttt{emproc} and \texttt{epproc}. Periods of high-background levels were assessed by extracting light curves in the 10--12~keV energy range. Given their higher sensitivity, we will only present the analysis of spectra obtained with the EPIC pn camera. The net exposure times and count rates of the observations are also listed in Table~\ref{tab:obs}. All EPIC pn background-subtracted spectra are presented in the top left panel of Fig.~\ref{fig:spec1} for the 0.3--10.0~keV energy range.

For discussion, we also extracted {\it XMM-Newton} Reflection Grating Spectrograph (RGS) spectra from the 2017 observations. The raw data were processed using the \texttt{rgsproc} task of {\sc sas}. Periods of high background were assessed inspecting that of CCD\#9 for the RGS1 and RGS2 instruments. The task \texttt{rgscombine} of {\sc sas} was used to combine the first-order RGS1 and RGS2 spectra to produce the spectrum shown in the top right panel of Fig.~\ref{fig:spec1}. We note that the RGS observations from other epochs did not result in acceptable signal-to-noise spectra.

\begin{figure*}
\begin{center}
\includegraphics[angle=0,width=0.95\linewidth]{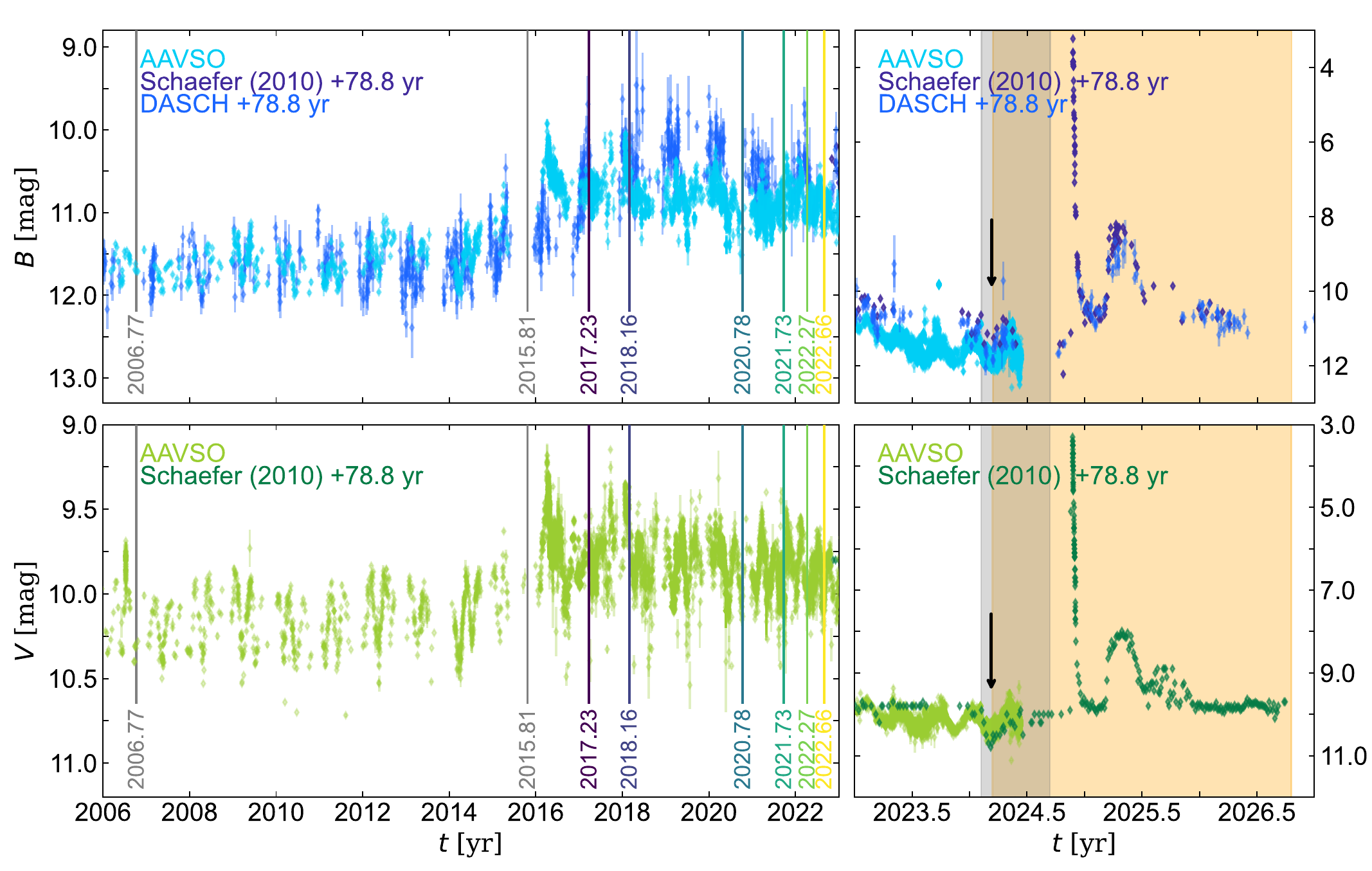} 
\caption{$B$- (top panels) and $V$-magnitude (bottom panels) light curves of T~CrB. Data from AAVSO, \citet{Schaefer2010} and DASCH are illustrated with different colours. The data from \citet{Schaefer2010} and DASCH have been displaced by 78.8 yr to fit the AAVSO light curve on year 2024.19. The vertical lines show the epoch of the X-ray observations analysed here. The grey- and yellow-shaded areas in the rightmost panels represents the possible next nova-like event predicted by \citet{Schaefer2023} and \citet{Munari2016}, respectively.}
\label{fig:lc}
\end{center}
\end{figure*}

\subsection{{\it NuSTAR} observations}

There are two different sets of archival {\it NuSTAR} observations of T~CrB. The first one was obtained on 2015-09-23T04:46:08 with a total exposure time of 79.8~ks and corresponds to the Obs.~ID. 30101046002. In addition, coordinated {\it NuSTAR} observations were obtained simultaneously with the 2020 {\it XMM-Newton} observations. These observations correspond to the Obs. ID. 80601307002 and were taken on 2020-09-08T00:36:09 with a total exposure time of 49.4~ks. Both sets of observations were retrieved from the HEASARC science archive\footnote{\url{https://heasarc.gsfc.nasa.gov/cgi-bin/W3Browse/w3browse.pl}}. The data were processed using the standard procedure to process {\it NuSTAR} observations using {\sc nupipeline} version 0.4.8. In both cases, we processed the two focal plane modules (A and B). The resultant 2015 and 2020 {\it NuSTAR} spectra (A and B) are presented in the bottom panels of Fig.~\ref{fig:spec1}.

\subsection{{\it Suzaku} observations}

Available {\it Suzaku} observations of T~CrB were also retrieved from the HEASARC science archive. These were obtained on 2006-09-06T22:44:21 with a total exposure time of 46.3~ks, corresponding to Obs.~ID. 401043010 and include observations performed with the XIS and HXD instruments. We extracted background subtracted spectra from the four XIS instruments (XIS0, XIS1, XIS2, and XIS3) following standard processing recipes\footnote{\url{https://heasarc.gsfc.nasa.gov/docs/suzaku/analysis/abc/node9.html}} using the {\sc HEASoft} package\footnote{\url{http://heasarc.gsfc.nasa.gov/ftools}} \citep{Heasoft2014}. The RMF and ARF files were created using the \texttt{xisrmfgen} and \texttt{xissimarfgen} {\sc HEASoft} tasks. Finally, spectra extracted from the XIS0, XIS2, and XIS3 were combined using the \texttt{addascaspec} task, originally designed for {\it ASCA} data but also helpful to combine spectra and responses of {\it Suzaku} XIS from the front-illuminated chips. The resultant {\it Suzaku} spectrum obtained from the back-illuminated chip is presented in the top left panel of Fig.~\ref{fig:spec1} alongside the {\it XMM-Newton} EPIC-pn data. This spectrum is the same as that analysed in \citet{Kennea2009}.

\subsection{Complementary observations}

We also retrieved the optical light curves of T~CrB from the American Association of Variable Star Observers (AAVSO)\footnote{\url{https://www.aavso.org/}} obtained with the optical $B$, $V$ and $I$ filters which are contemporaneous with the Observations obtained with the current generation of X-ray satellites. The data correspond to photometric measurements obtained between 2004 Mar 11 and 2024 June 11. The resultant light curves of the $B$ and $V$ filters are shown in Fig.~\ref{fig:lc} and a comparison between these two and the $I$ band photometry is presented in Appendix~\ref{sec:app}. This figure also shows the photometric measurements obtained from the Digital Access to a Sky Century @ Harvard (DASCH)\footnote{\url{http://dasch.rc.fas.harvard.edu/lightcurve.php}} and values listed in \citet{Schaefer2010}. The latter data sets were shifted +78.8 yr to make them coincident with the optical dip observed on the year 2024.19 (marked with an arrow in the right panels of Fig.~\ref{fig:lc}).

We note that this shift seems to suggests that T~CrB will experience its next nova-like event on 2024.9 (about the end of October of the present year). Other predictions suggest this event to take place on 2024.4$\pm$0.4 \citep{Schaefer2023} and 2025.5$\pm$1.3 \citep{Munari2016}. These two possible time ranges are illustrated in Fig.~\ref{fig:lc}.

\section{Data analysis and Results}
\label{sec:results}

The medium-resolution spectra presented in Fig.~\ref{fig:spec1} confirm that the 2006.77 epoch only exhibits the presence of a hard component ($E>2$ keV) with the clear presence of the Fe complex in the 6.0--7.0 keV energy range \citep{Luna2008}, corroborating the $\delta$-type classification of T~CrB, before the start of the active phase. The same plot also shows that, although exhibiting significant variability, all spectra extracted after 2017 are characterised by the presence of both soft and hard X-ray emission, hence motivating the later classification of T~CrB as a $\beta/\delta$ type X-ray-emitting symbiotic system.

At first look, data accumulated between 2017.24 and 2022.66 reveal a somewhat anti-correlation where the 0.3--2.0~keV band becomes weaker when the 2.0--10.0~keV intensifies \citep[see][for the discussion between the 2017.24 and 2018.16 epochs]{Zhekov2019}. However, this is not particularly true for the 2020.77 epoch. For example, the EPIC pn spectra of the 2020.77 and 2022.27 epochs show similar fluxes in the 2.0-10.0~keV energy range, but that of the 2020.77 epoch exhibits considerably less flux in the softer band (see top left panel of Fig.~\ref{fig:spec1}), similar to that of the 2022.66 epoch.

The {\it NuSTAR} spectrum obtained during 2015.81 epoch is consistent with the 2006.77 {\it Suzaku} data, which seems to suggest that the dramatic change in the hard X-ray emission (the appearance of the soft X-ray component) only started after the optical peak reached during 2016.35. A comparison between these two sets of observations is presented in the bottom left panel of Fig.~\ref{fig:spec1}. Consequently, these two epochs will be treated as a single one and a simultaneous joint model to these two epochs will help us assess the X-ray properties, up to energies of 50 keV, for times before T~CrB entered its current active phase. Evidently, this is also the case of the 2020.77 {\it XMM-Newton} observations and the 2020.77 {\it NuSTAR} data (see bottom right panel of Fig.~\ref{fig:spec1}).

It is interesting to note that the RGS spectrum obtained on 2017.24 shows the presence of emission lines (top right panel of Fig.~\ref{fig:spec1}). There seems to be emission from the C\,{\sc v}, S\,{\sc xi}, N\,{\sc vi}, Ar\,{\sc xv} and N\,{\sc vii} at 35.0~\AA,  31.5~\AA, 29.1~\AA, 27.05~\AA\, and 24.78~\AA, respectively, which have been also detected in other X-ray-emitting symbiotic systems \citep[see][]{Toala2023,Toala2023b}. 
Given that the RGS instruments are best suited to study the emission in the 0.33--2.5~keV energy range, the 2017.24 RGS spectrum is evidently tracing the contribution from the soft component. The estimated signal-to-noise of these emission lines range between 2.0 and 3.4 and their presence is also suggested by the medium-resolution EPIC MOS spectra (not shown here) but with evidently lower spectral quality. This suggests that the soft X-ray emission has a thermal origin \citep[see also the cases of R~Aqr and V694~Mon discussed;][]{Sacchi2024,Lucy2020}, instead of a black body as suggested in previous analysis of the {\it XMM-Newton} data \citep[see][]{Luna2018,Zhekov2019}.

\begin{figure}
\begin{center}
\includegraphics[angle=0,width=\linewidth]{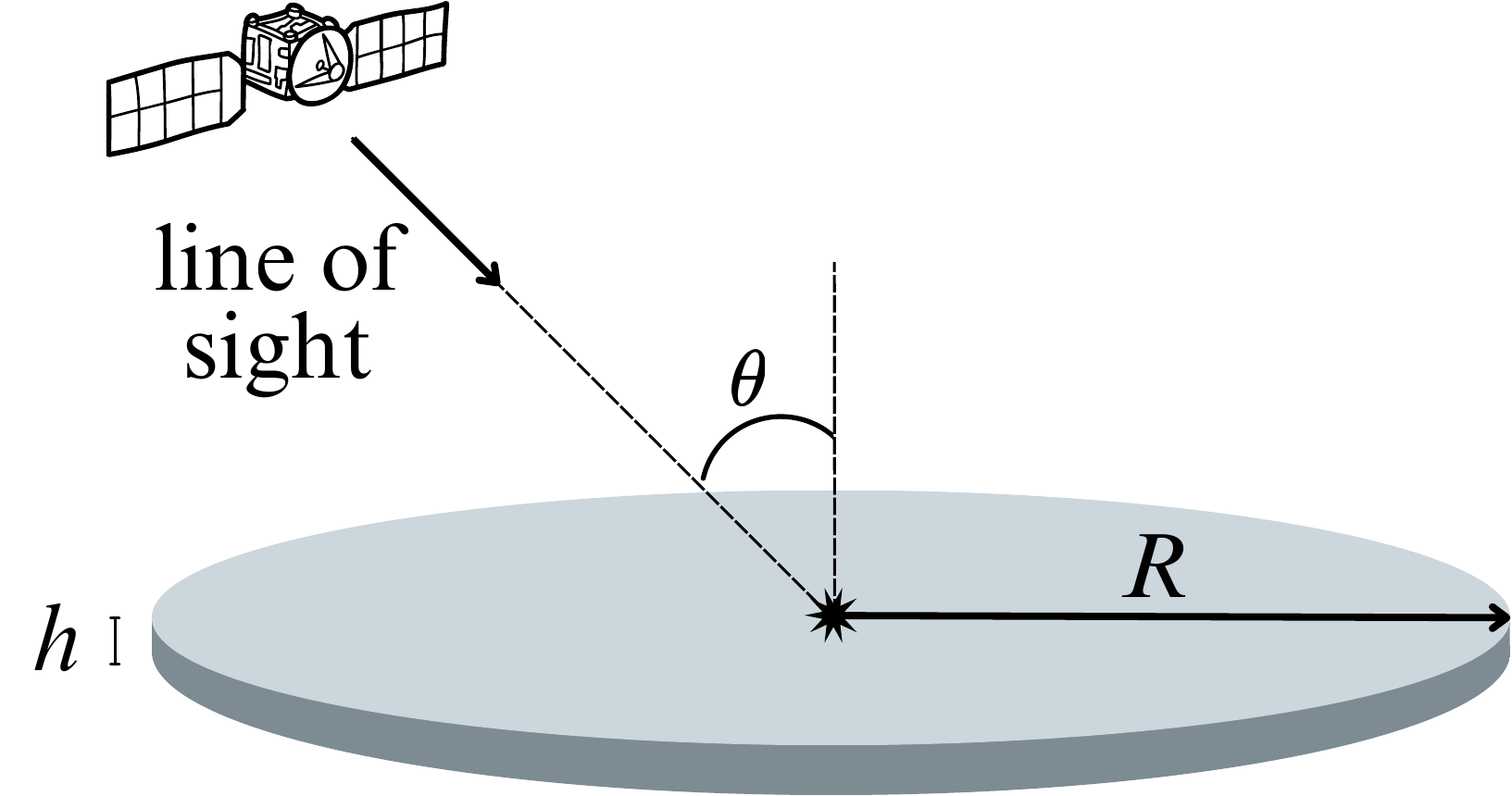}
\caption{A schematic view of the disk structure adopted defined for the \texttt{reflection} model produced with {\sc reflex}.}
\label{fig:disk}
\end{center}
\end{figure}

\begin{table*}
\begin{center}
\caption{Best-fit parameters of the spectral analysis of the X-ray observations of T~CrB. The observed ($f_\mathrm{X}$) and unabsorbed ($F_\mathrm{X}$) fluxes were computed for the 0.3--50.0~keV energy range. Boldface values represent the fixed values.}
\label{tab:analysis}
\setlength{\tabcolsep}{0.4\tabcolsep}  
\begin{tabular}{lcccccccc}
\hline
%&  & Epoch   &  &\\  
% & 2007.83 & 2013.37 & 2016.28 & 2020.31 
%\\
Parameter & Units & 2006.77$+$2015.81 & 2017.24 & 2018.16 & 2020.77 & 2021.73 & 2022.27 & 2022.66\\
\hline
Soft component \\
$N_\mathrm{H1}$  & 10$^{22}$ cm$^{-2}$   & {\bf 0.23} & 0.23$\pm$0.01    & {\bf 0.23}    & {\bf 0.23}    & {\bf 0.23}        & {\bf 0.23}     & {\bf 0.23} \\
$kT_1$           & keV                   & \dots         & 0.026$\pm$0.001 & 0.033$\pm$0.002 & 0.048$\pm$0.022 & 0.015$\pm$0.010 & 0.029$\pm$0.001 & 0.029$\pm$0.004 \\
$A_1$           & cm$^{-5}$              & \dots         & 262$\pm$94 & 5.8$\pm$2.1 & 0.011$\pm$0.004 & 0.010$\pm$0.004 & 16.3$\pm$5.8 & 2.1$\pm$0.8 \\
$f_\mathrm{X1}$ & erg~cm$^{-2}$~s$^{-1}$ & \dots         & 3.6$\times10^{-13}$&9.6$\times10^{-14}$& 5.1$\times10^{-15}$&4.2$\times10^{-15}$&7.4$\times10^{-14}$ & 8.8$\times10^{-15}$\\
$F_\mathrm{X1}$ & erg~cm$^{-2}$~s$^{-1}$ & \dots         & 3.2$\times10^{-11}$&6.6$\times10^{-12}$& 1.9$\times10^{-13}$&5.9$\times10^{-13}$&6.0$\times10^{-12}$ & 7.2$\times10^{-13}$\\
\hline
Hard component \\
$N_\mathrm{H2}$ & 10$^{22}$ cm$^{-2}$    & 25.3$\pm$0.9  & 59.4$\pm$7.0 & 38.1$\pm$6.8 & 27.9$\pm$1.7 & 45.3$\pm$1.7 & 29.1$\pm$1.8 & 33.1$\pm$1.2 \\ 
CF              &                        & 0.998$\pm$0.001 & {\bf 0.998} & {\bf 0.998} & {\bf 0.998} & {\bf 0.998} & {\bf 0.998} & {\bf 0.998} \\
$kT_2$          & keV                    & 14.8$\pm$1.6  & 2.8$\pm$0.8  & 5.9$\pm$1.6  & 7.7$\pm$0.5 & 9.5$\pm$0.6 & 7.7$\pm$0.7  & 8.7$\pm$0.5\\
$A_2$            & cm$^{-5}$             &  (1.7$\pm$0.1)$\times10^{-2}$ & (3.3$\pm$0.1)$\times10^{-3}$ & (2.2$\pm$0.1)$\times10^{-3}$ & (3.9$\pm$0.1)$\times10^{-3}$ & (7.4$\pm$0.2)$\times10^{-3}$& (3.7$\pm$0.1)$\times10^{-3}$ &  (9.4$\pm$0.3)$\times10^{-3}$\\
$f_\mathrm{X2}$ & erg~cm$^{-2}$~s$^{-1}$ &  2.7$\times10^{-11}$ & 3.8$\times10^{-13}$ & 1.1$\times10^{-12}$ & 3.3$\times10^{-12}$ & 4.7$\times10^{-12}$ & 2.6$\times10^{-12}$ & 6.7$\times10^{-12}$\\
$F_\mathrm{X2}$ & erg~cm$^{-2}$~s$^{-1}$ & 4.3$\times10^{-11}$ & 4.9$\times10^{-12}$ & 4.2$\times10^{-12}$ & 8.2$\times10^{-12}$ & 1.7$\times10^{-11}$ & 8.0$\times10^{-12}$ & 2.1$\times10^{-11}$\\
Reflection \\
$A_3$            & cm$^{-5}$             &  0.995$\pm$0.075 & 0.069$\pm$0.005 & 0.092$\pm$0.007 & 0.133$\pm$0.010 & 0.233$\pm$0.022 & 0.086$\pm$0.006 & 0.485$\pm$0.036 \\
$f_\mathrm{X3}$ & erg~cm$^{-2}$~s$^{-1}$ &  2.9$\times10^{-11}$ & 5.6$\times10^{-13}$ & 7.7$\times10^{-13}$ & 3.9$\times10^{-12}$ & 2.4$\times10^{-12}$ & 7.2$\times10^{-13}$ & 4.1$\times10^{-12}$\\
$F_\mathrm{X3}$ & erg~cm$^{-2}$~s$^{-1}$ & 2.9$\times10^{-11}$ & 6.1$\times10^{-13}$ & 8.1$\times10^{-13}$ & 4.0$\times10^{-12}$ & 2.6$\times10^{-12}$ & 7.6$\times10^{-13}$ & 4.3$\times10^{-12}$ \\
\hline
$f_\mathrm{X,TOT}$ & erg~cm$^{-2}$~s$^{-1}$& 5.6$\times10^{-11}$ & 1.3$\times10^{-12}$ & 2.0$\times10^{-12}$ & 7.2$\times10^{-12}$ & 7.1$\times10^{-12}$ & 3.4$\times10^{-12}$ & 1.1$\times10^{-11}$\\
$F_\mathrm{X,TOT}$ & erg~cm$^{-2}$~s$^{-1}$& 7.2$\times10^{-11}$ & 3.8$\times10^{-11}$ & 1.2$\times10^{-11}$ & 1.2$\times10^{-11}$ & 2.0$\times10^{-11}$ & 1.5$\times10^{-11}$ & 2.6$\times10^{-11}$\\
$L_\mathrm{X,TOT}$ & erg~s$^{-1}$      &  6.9$\times10^{33}$ &3.6$\times10^{33}$ & 1.2$\times10^{33}$ & 1.1$\times10^{33}$ & 1.9$\times10^{33}$ & 1.4$\times10^{33}$ & 2.5$\times10^{33}$ \\
\hline
$<F_{V}>$ & erg~cm$^{-2}$~s$^{-1}$& 6.7$\times10^{-10}$ & 1.1$\times10^{-9}$ & 1.4$\times10^{-9}$ & 9.3$\times10^{-10}$ & 9.6$\times10^{-10}$ & 9.3$\times10^{-10}$ & 9.1$\times10^{-10}$\\
$<F_{B}>$ & erg~cm$^{-2}$~s$^{-1}$& 2.6$\times10^{-10}$ & 6.0$\times10^{-10}$ & 7.6$\times10^{-10}$ & 4.6$\times10^{-10}$ & 5.8$\times10^{-10}$ & 5.0$\times10^{-10}$ & 5.3$\times10^{-10}$\\
$<F_{I}>$ & erg~cm$^{-2}$~s$^{-1}$& 4.0$\times10^{-9}$ & 4.8$\times10^{-9}$ & 5.2$\times10^{-9}$ & 4.5$\times10^{-9}$ & 4.2$\times10^{-9}$ & 4.4$\times10^{-9}$ & 4.0$\times10^{-9}$\\
\hline
$\dot{M}_\mathrm{acc}$ & M$_\odot$~yr$^{-1}$ & \dots & 2.4$\times10^{-8}$ & 4.1$\times10^{-8}$ & 2.0$\times10^{-8}$ & 2.4$\times10^{-8}$ & 2.3$\times10^{-8}$ & 2.4$\times10^{-8}$\\
\hline
\end{tabular}
\end{center}
\end{table*}

\begin{figure*}
\begin{center}
\includegraphics[angle=0,width=\linewidth]{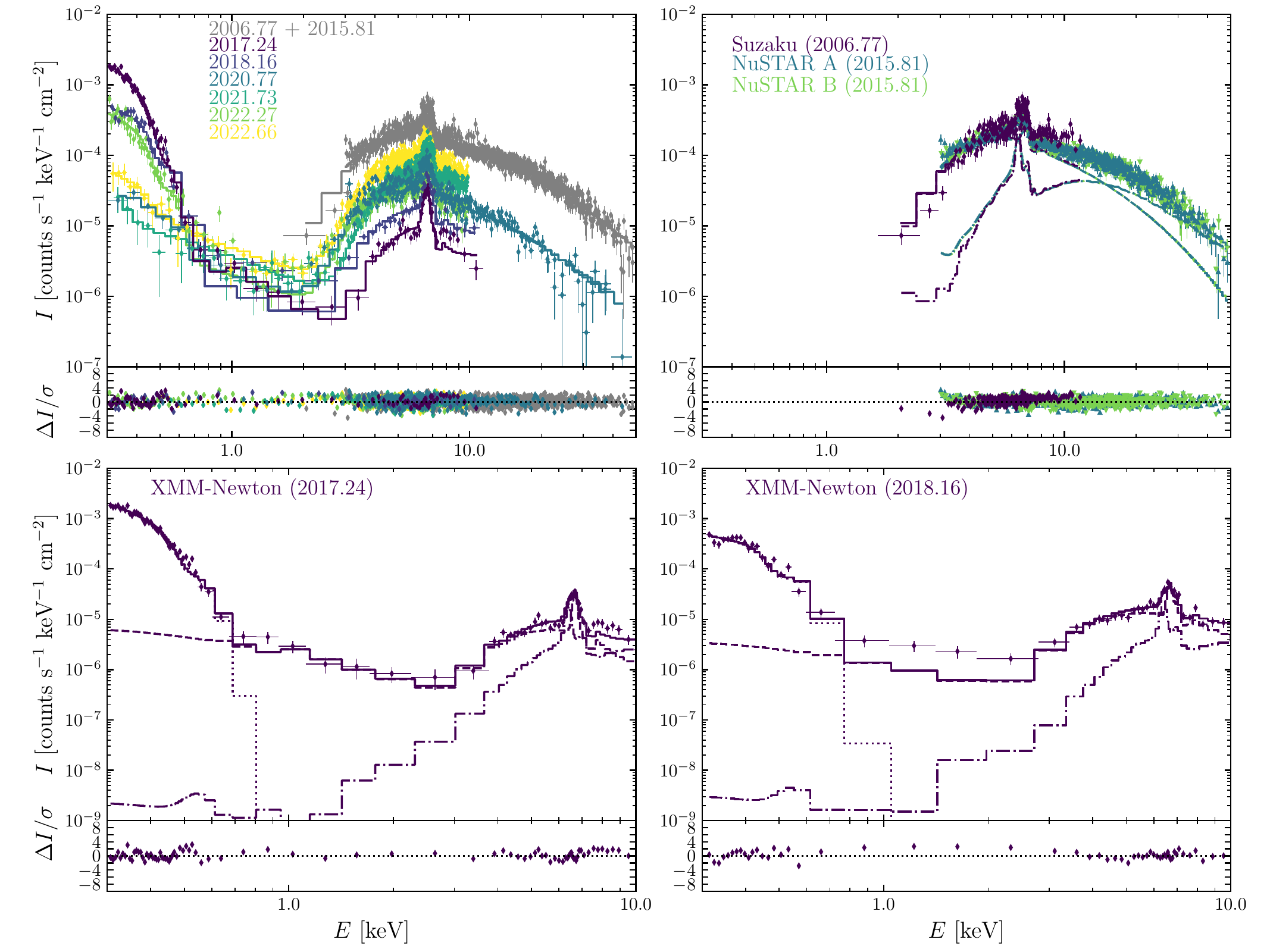}
\caption{Best-fit models (solid lines) obtained for the different epochs compared with their corresponding background-subtracted X-ray spectra. The dotted, dashed and dash-dotted lines represent the contributions from the soft plasma, heavily-extinguished plasma and the reflection component, respectively. See details in Table~\ref{tab:analysis}.}
\label{fig:spec2}
\end{center}
\end{figure*}

\subsection{Spectral analysis}

The spectral analysis of the X-ray data of T~CrB presented here was performed by means of the X-Ray Spectral Fitting Package \citep[{\sc xspec};][]{Arnaud1996}. Following previous analysis of the X-ray data of $\delta$ and $\beta/\delta$ symbiotic systems, including T~CrB \citep[e.g.,][]{Zhekov2019,Lucy2020}, we decided to model the heavily-extinguished hot component by adopting an optically-thin emission model \texttt{apec}, included in {\sc xspec}\footnote{\url{https://heasarc.gsfc.nasa.gov/xanadu/xspec/manual/node134.html}}. However, the presence of emission lines in the soft X-ray range unveiled by the 2017.24 data prevents us from modelling the soft range (0.3--2.0~keV) with a black body emission model. The soft X-ray emission was hence modelled with a second \texttt{apec} component. The extinction is taken into account by adopting photoelectric absorption using cross-sections of the \texttt{phabs} model, also part of {\sc xspec}.

All of the X-ray spectra of T~CrB analysed here show the presence of the unresolved Fe emission lines in the 6.0--7.0~keV energy range (see Fig.~\ref{fig:spec1}). The identification of the 6.4 keV Fe emission line in symbiotic systems is compelling evidence for the reflection mechanism generated by the inner regions of the accretion disk \citep{Eze2014,Ishida2009,Toala2024}. Employing a simplistic Gaussian component with no physical meaning behind it neglects the role of reflection and certainly diminishes the accuracy of the analysis.

Taking advantage of the analysis of the $\beta/\delta$ symbiotic system CH Cyg presented by our group \citep{Toala2023b}, we decided to use tailor-made reflection models in the spectral analysis of T~CrB. We produced 200 disk models using the radiative-transfer code {\sc reflex} \citep{Paltani2017} which simulates physical processes of X-ray photons propagating through matter around the central source, adopting Monte Carlo calculations to track individual photons. These models are characterised by an averaged column density $N_\mathrm{H,ref}$, a radius $R$, a thickness $h$ (such that $h=0.1R$) and a line of sight $\theta$. We sampled $N_\mathrm{H,ref}$=[10$^{23}$,10$^{24}$,10$^{25}$,10$^{26}$]~cm$^{-2}$ and $R$=[0.3, 0.5, 1.0, 10, 100]~AU. For each disk model, we created 10 spectral energy distributions (SED) adopting angles between 0 and 90$^{\circ}$ (see illustration in Fig.~\ref{fig:disk}). Finally, we converted these models to an additive one-parameter table (in fits format) associated with all the SEDs using the {\sc HEASoft} task \texttt{ftflx2tab}\footnote{\url{https://heasarc.gsfc.nasa.gov/lheasoft/help/ftflx2tab.html}}.

It is important to mention here that these disk structures do not represent only the accretion disk around the WD component, they are meant to represent all the structures in the circumstellar medium around the symbiotic system (including the accretion disk). For example, numerical simulations presented in \citet{Liu2017} and \citet{deValBorro2017} show complex toroidal-like circumstellar material around the accreting WD. However, for simplicity we will refer to this model as \texttt{reflection}, because is the morphological feature that produces the reflection, particularly, its innermost region (see discussion section).

We adopted the following model configuration:
\begin{equation}
    {\rm phabs}_1 \cdot {\rm vapec}_1 + {\rm CF} ({\rm phabs}_2 \cdot {\rm apec_2}) +  {\rm CF (phabs_1} \cdot {\rm phabs_2} \cdot {\rm reflection}),
\end{equation}
\noindent where CF is a covering factor which helps assessing the very likely porous (or nonuniform) distribution of material around symbiotic systems. Evidently, the analysis of the 2006.77+2015.81 epoch does not require the inclusion of the soft component. Our models adopt Solar abundances from \citet{Lodders2003}, but also tried models varying different abundances, in particular the Fe abundance, but these resulted in 1.1$\pm$0.1 times the Solar value.

A simultaneous fit to all spectra was produced leaving as free parameters the hydrogen column densities ($N_\mathrm{H1}$ and $N_\mathrm{H2}$) associated with the two \texttt{phabs} models, the two plasma temperatures ($kT_1$ y $kT_2$), the covering factor CF and the disk properties ($R$, $h$, $N_\mathrm{H,ref}$ and inclination with respect of the line of sight $\theta$).
The best model was selected by evaluating the reduced $\chi^2$ statistics ($\chi^2_\mathrm{DoF}$). This process required a few months of calculations given the large number of parameters in combinations with our \texttt{reflection} model.

\begin{figure*}
\begin{center}
\includegraphics[angle=0,width=\linewidth]{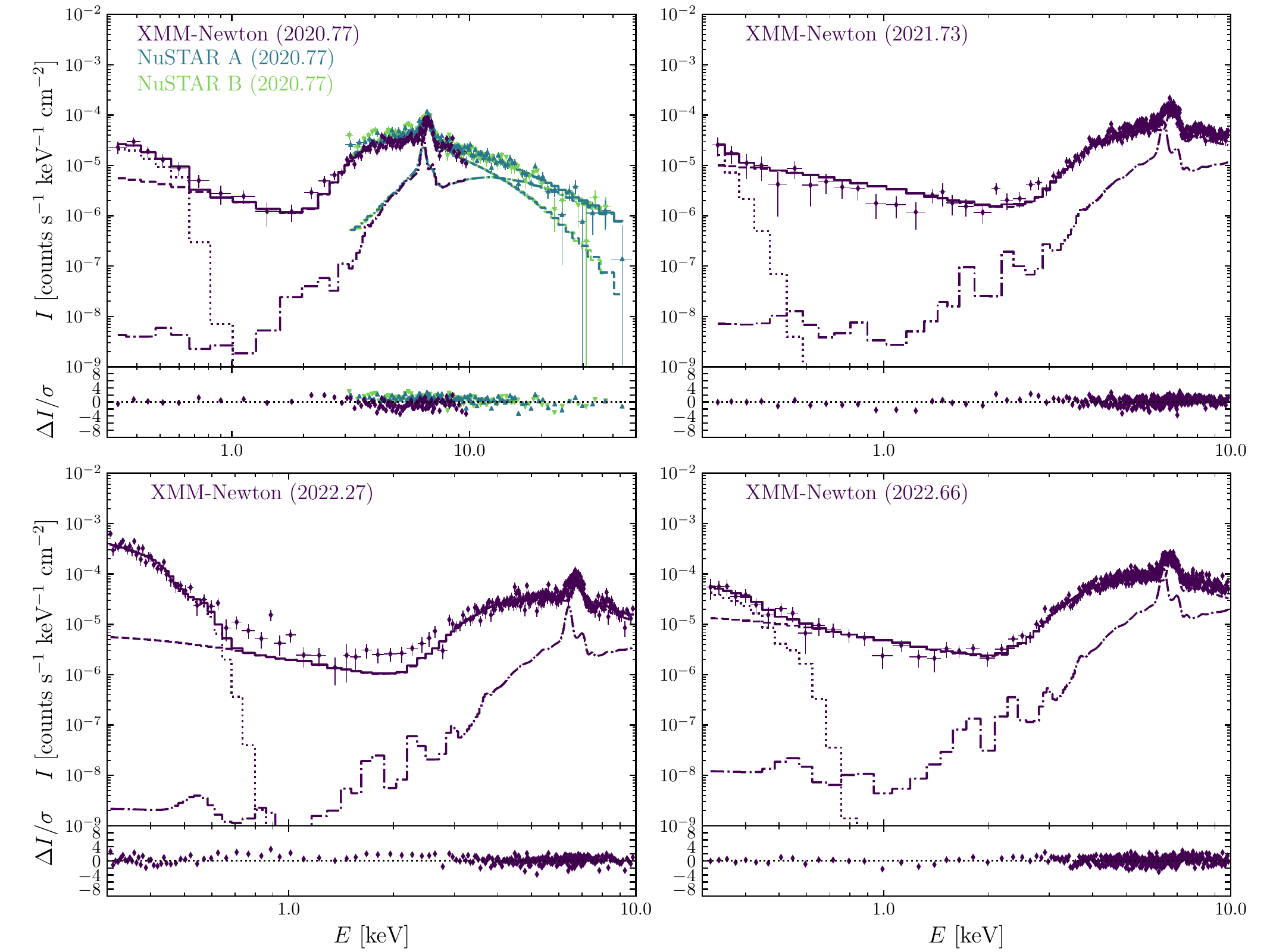}
\caption{Same as Fig.~\ref{fig:spec2} but for the 2020.77, 2021.73, 2022.27 and 2022.66 epochs. The dotted, dashed and dash-dotted lines represent the contributions of the soft plasma, heavily-extinguished plasma and the reflection component, respectively. See details in Table~\ref{tab:analysis}.}
\label{fig:spec3}
\end{center}
\end{figure*}

The best model ($\chi^{2}_\mathrm{DoF}$=3285.7/2639 = 1.25) resulted in the parameters listed in Table~\ref{tab:analysis}. The disk model that best reproduces the X-ray data is that with $R$=1 AU, $h$=0.1 AU, $N_\mathrm{H,ref}$=10$^{25}$~cm$^{-2}$ and an inclination of $\theta$=50$^{\circ}$. We note that larger disk models resulted in slightly worse fits because reflection is also produced in more external regions, while a disk of 0.5~AU or less produced a much worse fit. The reflecting disk structure can not be as small as 0.5~AU.
The contribution from the individual components are listed through their normalisation parameter $A$\footnote{The normalisation parameter in {\sc xspec} is defined as $A=10^{-14}\int n_\mathrm{e} n_\mathrm{H} dV / 4 \pi d^2$, with $n_\mathrm{e}$ and $n_\mathrm{H}$ as the electron and hydrogen number densities, $V$ is the volume, and $d$ is the distance.}, observed $f_\mathrm{X}$ and intrinsic $F_\mathrm{X}$ fluxes. The total fluxes and luminosities ($f_\mathrm{X,TOT}$, $F_\mathrm{X,TOT}$ and $L_\mathrm{X,TOT}$) for each epoch are also listed at the bottom part of this table. The best-fit models are compared with the observations in Fig.~\ref{fig:spec2} and \ref{fig:spec3}.

We used the 2017.24 epoch to restrict $N_\mathrm{H1}$ and used this as a fixed value for the rest of the epochs. Table~\ref{tab:analysis} shows that $N_\mathrm{H1}$ resulted in 2.3$\times10^{21}$~cm$^{-2}$, which is very close to that estimated by Galactic H\,{\sc i} column density towards T CrB \citep[see section 3.2 of][]{Zhekov2019}. The CF was left as a free parameter in the determination of the 2006.77+2015.81 epoch, which resulted in 0.998$\pm$0.001. This value was taken as a fixed parameter for the rest of the epochs. Fixed values are shown with boldface text in Table~\ref{tab:analysis}.

\begin{figure}
\begin{center}
\includegraphics[angle=0,width=\linewidth]{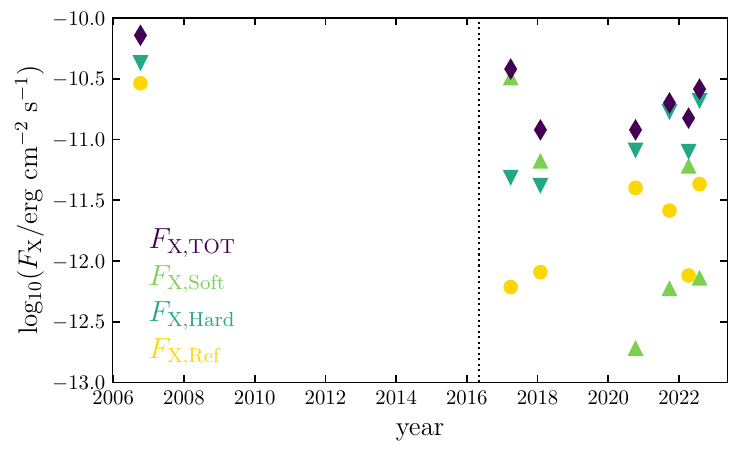}\\
\includegraphics[angle=0,width=\linewidth]{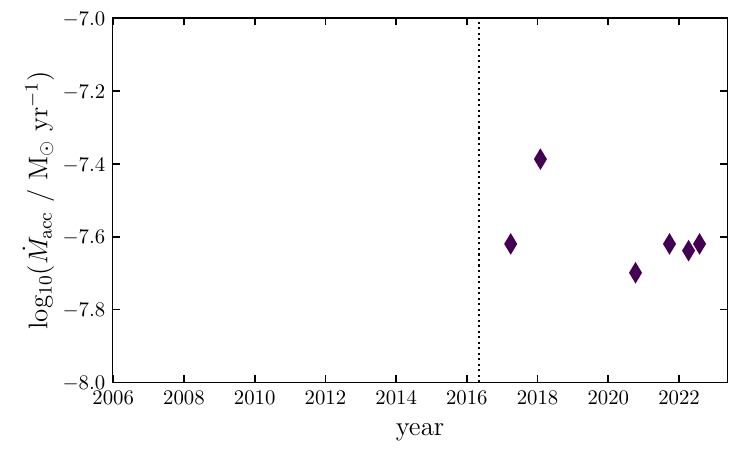}\\
\includegraphics[angle=0,width=\linewidth]{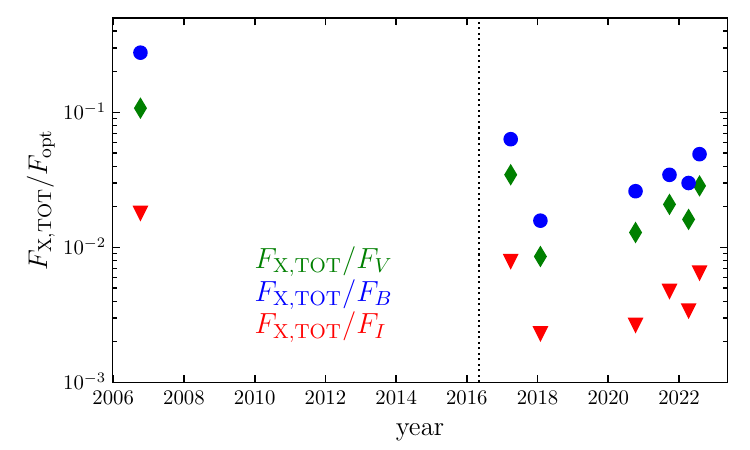}
\caption{Temporal evolution of the X-ray flux and its components (top panel), mass accretion rate $\dot{M}_\mathrm{acc}$ (middle panel) and the ratio of the X-ray flux over that obtained from the optical $B$, $V$ and $I$ bands (bottom panels). The vertical dotted line shows the time of the maximum optical peak reported on 2016.35.}
\label{fig:time_evol}
\end{center}
\end{figure}

The temperature of the soft component ($kT_1$) is more or less consistent within different epochs, with an averaged value of 0.03~keV (=3.5$\times10^{5}$ K), however, its observed flux varies about two orders of magnitudes between epochs, being that of 2017.24 the brightest soft component. The heavily-extinguished \texttt{apec} ($kT_2$) and the \texttt{reflection} components jointly fit the unresolved Fe feature and, in the case of the 2006.77+2015.81 and 2020.77 epochs, the energy range above 10 keV detected by {\it NuSTAR} (see top right panel of Fig.~\ref{fig:spec2} and top left panel of Fig.~\ref{fig:spec3}). The \texttt{reflection} model dominates for $E>$15 keV, leveraging the role of reflection in the hard energy range.
The heavily-extinguished thermal component shows a temperature ranged from $kT_2$=2.8~keV (=3.2$\times10^{7}$~K) to values of $kT_2$=14.8~keV (=1.7$\times10^{8}$~K) with an hydrogen column density varying from $N_\mathrm{H2}$=25$\times10^{22}$~cm$^{-2}$ to about 60$\times10^{22}$~cm$^{-2}$. The top panel of Fig.~\ref{fig:time_evol} illustrates the variation of the total intrinsic flux $F_\mathrm{X,TOT}$ and the corresponding contributions from the three components ($F_\mathrm{X,soft}$, $F_\mathrm{X,Hard}$ and $F_\mathrm{X,Ref}$).

It is interesting to note that according to the best-fit model, the \texttt{reflection} component does not dominate the 2.0--4.0 keV energy range, but instead there seems to be a competition between the hard X-ray-emitting plasma component, the reflection and its inclination angle. This situation seems to be the result of the inclusion of the CF \citep[see its discussion in][]{Zhekov2019}. In this case, not all the emission is absorbed by the hydrogen column density producing the low-energy tail of the hard X-ray plasma component (see Fig.~\ref{fig:spec2} and \ref{fig:spec3}).  In order to investigate the dependence of the resulting fit on the CF, we also performed spectral fits by removing the CF, which resulted in a bit worse quality model ($\chi^{2}_\mathrm{DoF} \approx 1.4$). To improve those models we would need an extra plasma component to fit the 0.9-2.0~keV energy range, but that would increase the number of free parameters in our study. A better understanding on this situation will be assess by the inclusion of density distributions around symbiotic systems resulting from detailed hydrodynamic simulations. Something that will be pursuit in a subsequent study but is out of the scope of the present paper.

\section{Discussion}
\label{sec:discussion}

As mentioned above, using the 2006.77+2015.81 spectrum it was possible to assign a $\delta$-type to T~CrB \citep[see table 1 in][]{Merc2019}, but any other latter medium-resolution X-ray spectrum can undoubtedly be used to classify this system as a $\beta/\delta$-type. This strongly suggest that the $\alpha$, $\beta$, $\delta$, and $\beta/\delta$ classification scheme, initially proposed by \citet{Murset1997} and expanded by \citet{Luna2013}, is not fundamental. In contrast, it might be describing a specific time in the ever-changing state of X-ray-emitting symbiotic systems. The variations in the spectral properties within different epochs only mimic the extremely variable nature of these systems. The multi-epoch, multi-instrument observational data collected for T~CrB allow us to peer into the dramatic changes that a symbiotic system experiences by evolving from a steady-state phase to an active one.

\subsection{The disk properties}

Within the assumptions of the analysis presented here, the proposed disk model can reproduce satisfactorily the different epochs of observations. The best-fit is achieved with an inclination angle between the symmetry axis of the disk and the line of sight of $\theta=50^\circ$, which is very close to the value of 60$\pm$5$^{\circ}$ reported by \citet{Belczynski1998}. 

\citet{Luna2018} estimated a radius for the accretion disk in the steady state of 0.75~R$_\odot$(=3.5$\times10^{-3}$~AU), however, a very recent optical monitoring of T~CrB
presented by  \citet{Zamanov2024} concluded that its accretion disk has a size that ranges between 60~R$_\odot$(=0.28~AU) and 120~R$_\odot$(=0.56~AU), with an averaged value of 89$\pm$19~R$_\odot$(=0.42$\pm$0.09~AU). This value is about the same expected value as that of the Roche lobe of the WD component in T~CrB. Our best fit model requires the disk structure to have a value twice as that estimated by \citet{Zamanov2024}, $R$=1~AU. Which is not dramatically large if one takes into account the actual density distribution around the accreting WD in a symbiotic system as predicted by numerical simulations \citep[see for example fig.~2 in][]{Lee2022}. Again, we remark that smaller disk structures resulted in unacceptable models because they do not produce the necessary reflection component to produce good fits.

In fact, individual epochs might likely require different disk structures with slightly different parameters as demonstrated by \citet{Zamanov2024}. This can be particularly seen for the 2018.16 and 2022.27 epochs which have some difficulties fitting the 0.9--3.0 keV energy range. 
However, a more realistic accretion disk model, with an intricate 3D spiral structure as that found by hydrodynamical simulations \citep[e.g.,][]{deValBorro2017,Liu2017,Lee2022}, might be needed to reproduce the fine details of the observations (a situation that is also suggested by the CF parameter). Still, the analysis presented here demonstrates that the contribution from the reflection component should not be easily discarded as it significantly contributes to the total X-ray flux for some of the epochs and helps fitting the Fe complex and dominates the spectrum for energies above 15~keV. Furthermore, in some cases, the reflection component has a larger contributing flux ($F_\mathrm{X,Ref}$) than other components (see Table~\ref{tab:analysis} and the top panel of Fig.~\ref{fig:time_evol}).

\subsection{On the presence of a bipolar ejection}

In general, the analysis of the X-ray spectra from the 2006.77+2015.81, 2017.24 and 2018.16 epochs resulted in properties that are very similar to those reported by previous authors \citep{Luna2008,Luna2018,Kennea2009,Zhekov2019}. The heavily extinguished ($N_\mathrm{H2}>10^{23}$~cm$^{-2}$) component $kT_2$, very likely produced at the boundary layer, has a larger temperature for the steady-state epoch (2006.77+2015.81) and then is reduced after the start of the active phase. The rise of the soft component is coincident with the beginning of the active phase. Nevertheless, as previously stated, the rise of the soft component is not always coupled with a decay of the hard one.

One of the major differences between the present work and previous ones is the interpretation of the soft component. The presence of emission lines in the 2017.24 {\it XMM-Newton} spectra prevents from adopting a black body model and, instead, this emission is modelled with a thermal plasma component ($kT_1$). The best model suggests that the soft energy range can be modelled by plasma with temperatures between 0.015~keV (=1.7$\times10^{5}$~K) and 0.048~keV (=9.7$\times10^{5}$~K), which are very similar to those obtained for $\alpha$-type symbiotic systems that are typically attributed to thermonuclear burning on the surface of the WD component \citep[e.g.,][]{Orio2007}. However, in such cases the estimated luminosities are extremely high, up to 6 orders of magnitude larger.

Assuming that the soft temperatures in T~CrB are produced by a strong adiabatic shock, where the plasma temperature is a function of the gas velocity \citep[$T\propto\varv^2$;][]{Dyson1997}, the estimated temperatures can be translated into gas velocities of 110--200 km~s$^{-1}$. If this is to be the case, it is very likely that the soft X-ray emission is produced by the presence of bipolar ejections (or jets) emanating from the binary system, carving their way farther away from T~CrB. A situation that has been found in other $\beta/\delta$ systems, for example in the case of the best resolved symbiotic system, R~Aqr \citep{Kellogg2001,Kellogg2007,Toala2022,Sacchi2024}, and that of HM~Sge  and V694~Mon \citep[see][and references therein]{Toala2023,Corradi1999,Lucy2020}. This would help explain the fact that the soft component has a hydrogen column density two orders of magnitude smaller than that of the heavily-extinguished hot component ($N_\mathrm{H1} < N_\mathrm{H2}$), more consistent with the Galactic contribution. The model resulted in a slightly different $kT_1$ value for each epoch which suggests a variable nature of the velocity of the possible jet in T~CrB.

\subsection{Consequences for the temporal evolution of T~CrB}

The temperature of the heavily-extinguished plasma ($kT_2$) is estimated to be 14.8$\pm$1.6 keV in the steady-state epoch (2006.77$+$2015.81), with properties very similar to those estimated by previous authors \citep[][]{Luna2008,Kennea2009}. After the increase in the optical activity in 2016.35, the analysis of the X-ray observations suggests that $kT_2$ has decreased, going to 2.8$\pm$0.8 keV in the 2017.24 epoch to an averaged value of $\approx$8.0~keV in the more recent years (2018.16--2022.66). 

Changes in the temperature of the boundary layer can be used to assess $\dot{M}_\mathrm{acc}$ onto the WD component. Theoretical estimations have shown that high plasma temperatures are expected for low $\dot{M}_\mathrm{acc}$ and vice versa \citep[e.g.,][]{Pringle1979,Patterson1985}. In particular, \citet{Patterson1985} used a simple accretion disk model and found that low $\dot{M}_\mathrm{acc}$ makes the boundary layer to have low density and low optical depth, which does not cool efficiently and remains at high temperatures. On the other hand, high $\dot{M}_\mathrm{acc}$ produce a denser boundary layer that becomes optically thicker, cooling down and radiating its energy at lower temperatures. 
{\bf \citet{Patterson1985}} proposed that $\dot{M}_\mathrm{acc}$ and the plasma temperature of the boundary layer (in our case $kT_2$) are related through the equation
\begin{equation}
    kT_2 \approx 1.3 \left(\frac{M_\mathrm{WD}}{0.7~\mathrm{M}_\odot}\right)^{3.6} \left( \frac{10^{16}~\mathrm{g}~\mathrm{s}^{-1}}{\dot{M}_\mathrm{acc}} \right),
    \label{eq:acc1}
\end{equation}
\noindent with $M_\mathrm{WD}$ as the mass of the WD. Estimated to be 1.32 M$_\odot$ in the case of T~CrB \citep{Shara2018}.

Using Eq.~(\ref{eq:acc1}) we estimate that $\dot{M}_\mathrm{acc}$ is lower in the steady-state phase (2006.77+2015.81 epoch) with a value of 1.4$\times10^{-10}$~M$_\odot$~yr$^{-1}$ and then increases after the start of the active phase to an averaged  value of
3.4$\times10^{-10}$ for the 2017.24--2022.66 epochs. However, we note that previous $\dot{M}_\mathrm{acc}$ estimations for T~CrB resulted in higher values, particularly reaching to values $\gtrsim10^{-8}$~M$_\odot$~yr$^{-1}$ \citep[see, e.g., table 1 in][]{Zamanov2023}.

On the other hand, we can estimate $\dot{M}_\mathrm{acc}$ by assuming that the accretion process produces a certain bolometric luminosity $L_\mathrm{acc}$ such as 
\begin{equation}
    L_\mathrm{acc} =  \frac{G M_\mathrm{WD}}{\cos \theta~ R_\mathrm{WD}} \frac{\dot{M}_\mathrm{acc}}{2},
\label{eq:acc2}
\end{equation}
\noindent were $G$ as the gravitational constant and $R_\mathrm{WD}$ is the radius of the WD component. By assuming that the signature of accretion is the X-ray luminosity of the boundary layer (in our case that corresponding to $F_\mathrm{X2}$), we obtained $M_\mathrm{acc}$ values of a few times $10^{-9}$, where the largest value is that of the steady state phase and still over an order of magnitude below previous predictions.

To be able to use Eq.~(\ref{eq:acc2}) we need to account for all the luminosity produced by accretion in the different bands. As a lower limit, we can approximate $L_\mathrm{acc}$ only by accounting the luminosities obtained in X-rays and the optical bands. That is, by simply assuming $L_\mathrm{acc}\approx L_\mathrm{X} + L_\mathrm{opt}$. Taking advantage of the $L_\mathrm{opt}$ estimations of the accretion disk presented in \citet{Zamanov2023} for the active phase of T~CrB we estimate $\dot{M}_\mathrm{acc}$ values of 2.4$\times10^{-8}$, 4.1$\times10^{-8}$, 2.0$\times10^{-8}$, 2.4$\times10^{-8}$, 2.3$\times10^{-8}$ and 2.4$\times10^{-8}$ for the 2017.24, 2018.16, 2020.77, 2021.73, 2022.72 and 2022.66 epochs. These $\dot{M}_\mathrm{acc}$ estimates are also listed in the bottom row of Table~\ref{tab:analysis} and are further illustrated in the middle panel of Fig.~\ref{fig:time_evol}. It is evidently expected that the mass accretion rate to be smaller for the pre-active phase.

Given that the X-ray luminosity is about $\approx$1--10 per cent of $L_\mathrm{opt}$, our $\dot{M}_\mathrm{acc}$ estimations only represent slightly improved values than those estimated by \citet{Zamanov2023}. We particularly note the case of the 2017.24 epoch that has  contemporary optical observations in \citet{Zamanov2023}. These authors estimate $\dot{M}_\mathrm{acc}$=1.8$\times10^{-8}$~M$_\odot$~yr$^{-1}$ and accounting for the X-ray emission this is estimated to be about 30 per cent larger. These results illustrate that applying Eq.~(\ref{eq:acc2}) by only adopting the X-ray luminosity of the boundary layer one would need to apply efficiency factors $\eta \approx$20--100 to recover more realistic $\dot{M}_\mathrm{acc}$ values.

To explain the different observational properties of T~CrB, we propose the following scenario. T~CrB had a steady-state phase before the onset of the current active phase that started sometime around 2015. However, the X-ray properties did not change by 2015.81 and extreme changes were very likely produced after the peak in optical observations, around 2016.35. As reported by previous works, the active X-ray phase of T~CrB could be attributed to the accretion disk instabilities \citep[e.g.,][]{Luna2018}. Material from the disk enters the boundary layer, naturally increasing the amount of material that is accreted onto the WD component ($\dot{M}_\mathrm{acc}$ increases). The density increases as well as its optical depth, causing a reduction of the plasma temperature (see above). The accretion process onto the WD might not be efficient and, consequently, a jet-like ejection is produced creating adiabatically-shocked hot material towards the polar directions of the orbital plane producing the soft spectral component.

\begin{figure*}
\begin{center}
\includegraphics[angle=0,width=0.9\linewidth]{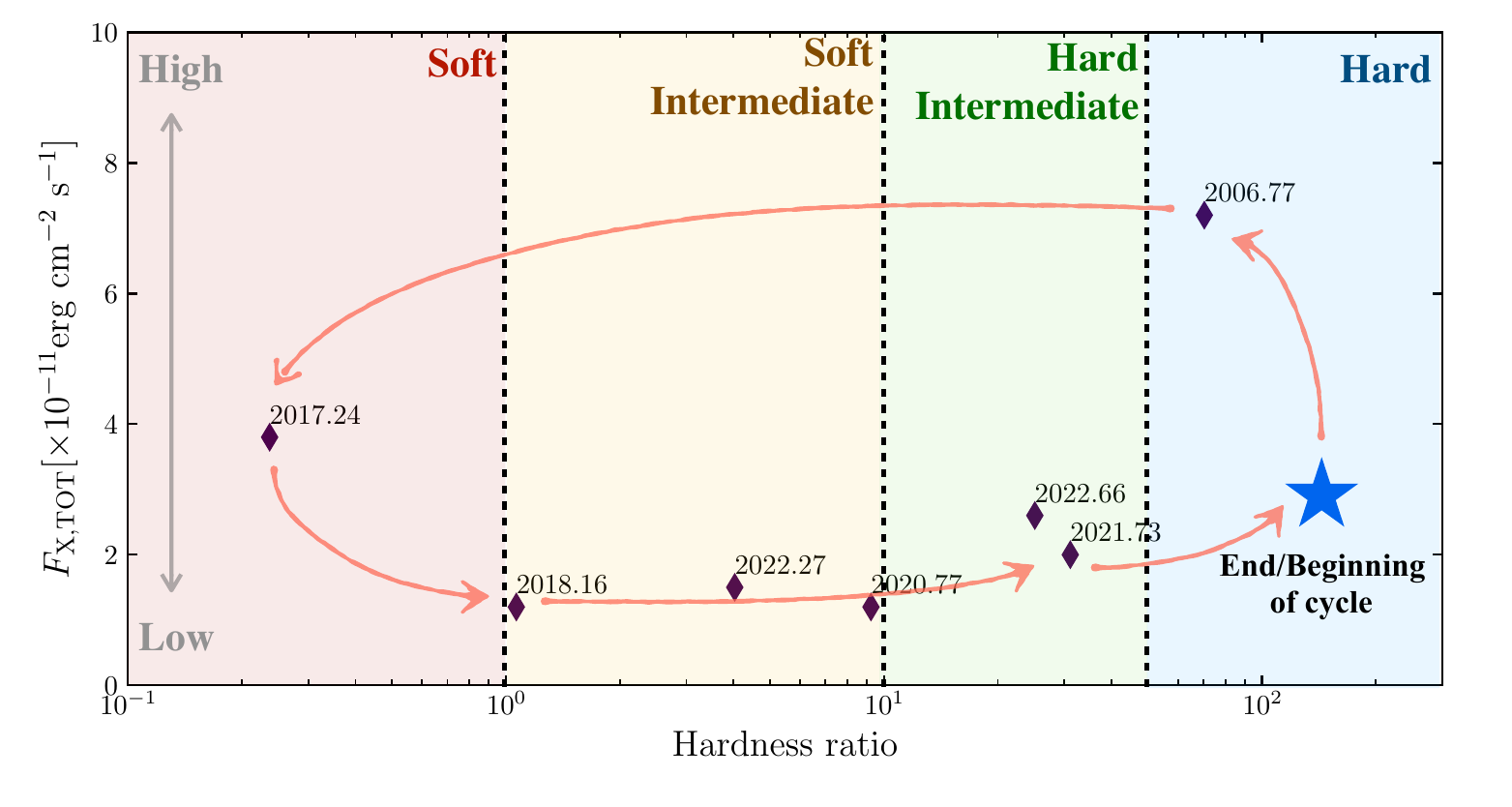}
\caption{Hardness-Intensity diagram (HID) of the evolution of T~CrB. Arrows show the evolutionary path of the X-ray properties of T~CrB (see text for details). Coloured, shaded areas show the different regimes during the evolution of T~CrB. The beginning and end of the cycle are marked with a coincident (blue) star. The hardness ratio is defined as the ratio of the observed [3.0--10.0 keV] band over the [0.3--3.0 keV].}
\label{fig:hardness}
\end{center}
\end{figure*}

The increase in the accretion rate of material onto the WD is somehow inversely proportional to the total X-ray flux. There is a clear anti-correlation between the total X-ray flux $F_\mathrm{X,TOT}$ and the optical measurements. The top panel of Fig.~\ref{fig:time_evol} shows that $F_\mathrm{X,TOT}$ was higher by the 2006.77 epoch and experienced a reduction already shown by 2017.24. A minimum in $F_\mathrm{X,TOT}$ was reached between the 2018.16 and 2020.77 epochs and started recovering in the later epochs of observation.

A reduction of the ratio between the X-ray emission and that of the optical band ($F_\mathrm{X}/F_\mathrm{opt}$) is expected for larger $\dot{M}_\mathrm{acc}$ \citep[see, e.g., fig. 7 of][]{Patterson1985}. Fig.~\ref{fig:time_evol} bottom panel shows the evolution with time of $F_\mathrm{X}/F_\mathrm{opt}$ for the $B$, $V$ and $I$ bands which can be directly compared with the evolution of the $\dot{M}_\mathrm{acc}$ in the middle panel. To calculate optical fluxes and luminosities from the AAVSO $I$, $V$ and $B$ photometric observations, we estimated median values for time intervals of half a year around the same dates as the X-ray observations. These fluxes ($<F_{I}>$, $<F_{V}>$ and $<F_{B}>$), calculated for each epoch are also listed at the bottom of Table~\ref{tab:analysis}.

The current analysis of the multi-epoch observations presented here shows that the contribution from the different spectral components (soft, hard and that of the reflection) do not follow a simple trend (see top left Fig.\ref{fig:time_evol}). The contribution from the hot plasma in the boundary layer, the plasma temperature of the jet-heated material, and the reflection contribution exhibit an apparent stochastic variability in time scales of years. But the variation of the X-ray properties of T~CrB describes a similar pattern of evolution as other accretion-powered systems (black hole X-ray binaries, neutron stars and AGN) when presented in the {\it hardness-intensity diagram} \citep[HID;][]{Belloni1997,Belloni2016,MD2014,Svoboda2017}.

Fig.~\ref{fig:hardness} shows the HID of T~CrB where we compare the total X-ray flux $F_\mathrm{X,TOT}$ versus the observed hardness ratio defined as the ratio of the 3.0--10.0~keV band over the 0.3--3.0~keV band. T CrB has evolved similarly to that described by, for example, \citet[][]{Motta2021} for black hole X-ray binaries. The evolution starts and finishes at the bottom rightmost part of the diagram (Low Hard) where, evidently, we do not have any X-ray observation, but is the region where the latest observations are converging to. This is denoted by a (blue) star in Fig.~\ref{fig:hardness}. It seems like, after the nova-like event experienced during 1946.1, T~CrB entered the steady-state phase represented by the 2006.77 epoch going into the High Hard region of the HID. An evolutionary path that is expected to repeat after the next nova-like event predicted to occur in the near future. Very likely after the outburst experienced in 2016.35, this symbiotic system moved to the Soft part of the diagram illustrated by the 2017.24 epoch. The subsequent evolution moved T~CrB into the Low part of the Soft and Hard intermediate phases. Vertical dashed lines are shown in the figure for illustration purposes separating the 2018.16, 2020.77 and 2022.27 epochs, where the soft X-ray component is still considerably, from those of 2021.73 and 2022.66 where the soft X-ray emission has almost vanished (see Fig.~\ref{fig:spec2} and \ref{fig:spec3}). We note that T~CrB went back and forth between the Soft Intermediate and Hard Intermediate phases between 2021.73 and 2022.66, a characteristic that is not unusual in other compact accreting systems \citep[see][and references therein]{Motta2021}.

T~CrB is practically the first symbiotic system with such detailed monitoring that allowed us to trace an almost complete loop in the HID, going from the steady-state phase into just a couple of years before its next nova-like event. This opens a new window for the studies of accreting WDs in symbiotic systems, adding to the multi-scale reach of accretion physics around compact objects. It would be interesting to corroborate that such evolution in the HID is the same for other symbiotic recurrent nova systems (for example, RS Oph) or any other accreting WD in a binary system.

\section{Summary}
\label{sec:summary}

We presented the analysis of archival X-ray observations of the symbiotic recurrent nova system T~CrB obtained with {\it NuSTAR}, {\it Suzaku} and {\it XMM-Newton} covering the 2006.77--2022.66 period. The analysis of the X-ray spectra was performed by including a reflection model produced by the presence of the accretion disk. The combination of these data in addition to optical monitoring allowed us to study the variable behaviour of T~CrB in a time span of about 16 years. Our main findings can be summarised as follows:
\begin{itemize}
    \item The multi-epoch spectra presented here confirm the extremely variable nature of the X-ray emission in T~CrB. Even within the observations covering the current active phase, there is no clear trend between the flux of the soft X-ray range ($E=$0.3--2.0~keV) and the hard ($E>$2.0~keV) range.

    \item The spectral analysis suggests that a disk structure with a radius of $R=$1 AU, a width of $h$=0.1 AU, averaged column density $N_\mathrm{H,Ref}=10^{25}$~cm$^{-2}$ and an inclination of $\theta$=50$^{\circ}$ between the light of sight and its symmetry axis can explain all of the X-ray spectra, naturally reproducing the 6.4~keV Fe emission line. The reflection component is necessary to explain the emission in the 10.0--50.0~keV spectral range detected in the {\it NuSTAR} data, dominating the $E>15$~keV energy range. The proposed reflecting disk structure is of the same order of the Roche lobe radius estimated for the WD component, only a factor of $\sim$2 larger than the sizes estimated by \citet{Zamanov2024}. Our disk model seem to suggest that not only the accretion disk produces reflection, but some of this effect might be also produced by the high-density material in the vicinity of the accretion disk.

    \item The presence of emission lines in the 2017.24 {\it XMM-Newton} data prevents from adopting a black body emission model to fit the soft spectral range. Instead, we use optically-thin emission plasma models that resulted in an average temperature of $kT_1$=3.5$\times10^{5}$~K. Although these temperatures are also observed in $\alpha$-type symbiotic systems their luminosities are orders of magnitude larger. We then suggest that these plasma temperatures are best explained by the presence of variable jet-like emissions with velocity values ranging from 110 to 200 km~s$^{-1}$, an explanation already invoked to explain the softest emission in other $\beta/\delta$ symbiotic systems (e.g., R~Aqr, HM Sge and V694 Mon).

    \item As expected from theoretical predictions, the evolution of the plasma temperature of the boundary layer ($kT_2$) is consistent with the increase of the activity in T~CrB. A high plasma temperature of 14.8~keV is observed in epochs before the onset of the current active phase. Conversely, lower plasma temperatures in the current active phase ($kT_2$=2.8--9.5~keV). 
    
    \item Using optical and X-ray flux estimates, we slightly improved previous the estimations for the mass accretion rate $\dot{M}_\mathrm{acc}$. An averaged value of 2.6$\times10^{-8}$~M$_\odot$~yr$^{-1}$ for the current active phase.
    
    \item The global evolution with time of the X-ray flux is anti-correlated with the optical properties of T~CrB as also found in other sources and predicted by the disk model, where higher accretion rates are correlated with lower ratios of the X-ray flux over that obtained from optical observations. We proposed a scenario in which T~CrB evolved from a quiescent, steady-state phase with high temperature into the current active stage by (very likely) the production of disk instabilities. In the current phase, $\dot{M}_\mathrm{acc}$ increased making the boundary layer material optically thicker and, consequently, radiating at lower temperatures. We suggest that the accretion process is not efficient enough and produces mas ejections towards the polar areas of the accretion disk (jets) that shock the circumstellar medium around T~CrB with velocities around 110--200~km~s$^{-1}$ producing the soft X-ray emission.

    \item T~CrB exhibits a loop evolution in the hardness-intensity diagram (HID), very similar to what has been observed in black hole binaries, accreting neutron stars and AGN. This parallelism pushes forward the idea that accreting WDs within symbiotic systems are also windows into the multi-scale reach of accretion physics around compact objects. The natural next step is to corroborate that this is also the case for other symbiotic systems. 
        
\end{itemize}

T~CrB is a unique case of a symbiotic recurrent nova system for several reasons. It is one of the few symbiotic stars with available multi-epoch, multi-instrument X-ray observations that have observed the steady-state phase and its evolution through the active phase. T~CrB is of particular interest given that it is predicted to experience its next nova-like outburst in the near future.

\section*{Acknowledgements} 

The authors are in debt to an anonymous referee for comments and suggestions that helped improve the presentation and discussion of the results. J.A.T. and D.A.V.T. acknowledge support from the UNAM PAPIIT project IN102324. O.G.M. thanks Fundaci\'{o}n Marcos Moshinsky (Mexico), UNAM PAPIIT project IN109123 and CONAHCyT "Frontera de la Ciencia" project CF-2023-G100. JAT thanks Emilio Tejeda for helping with the sketch of the disk model presented here and for long discussion sessions that helped draft the first version of this manuscript. D.A.V.T. thanks Consejo Nacional de Humanidades, Cientica y Tecnolog\'{i}a (CONAHCyT, Mexico) for a student grant. This work is based on observations obtained with {\it XMM-Newton}, an European Science Agency (ESA) science mission with instruments and contributions directly funded by ESA Member States and NASA. This research also made use of data obtained from the {\it Suzaku} satellite, a collaborative mission between the space agencies of Japan (JAXA) and the USA (NASA). This research has made use of data from the {\it NuSTAR} mission, a project led by the California Institute of Technology, managed by the Jet Propulsion Laboratory, and funded by the National Aeronautics and Space Administration. We acknowledge with thanks the variable star observations from the AAVSO International Database contributed by observers worldwide and used in this research. This work has made extensive use of NASA's Astrophysics Data System.

%%%%%%%%%%%%%%%%%%%%%%%%%%%%%%%%%%%%%%%%%%%%%%%%%%



%\bibliographystyle{mnras}

\begin{thebibliography}{99}


\bibitem[Arnaud(1996)]{Arnaud1996} Arnaud, K.~A.\ 1996, Astronomical Data Analysis Software and Systems V, 101, 17

\bibitem[Bailer-Jones et al.(2021)]{BailerJones2021} Bailer-Jones, C.~A.~L., Rybizki, J., Fouesneau, M., et al.\ 2021, \aj, 161, 147. doi:10.3847/1538-3881/abd806

\bibitem[Belczynski \& Mikolajewska(1998)]{Belczynski1998} Belczynski, K. \& Mikolajewska, J.\ 1998, \mnras, 296, 77

\bibitem[Belloni et al.(1997)]{Belloni1997} Belloni, T., M{\'e}ndez, M., King, A.~R., et al.\ 1997, \apjl, 479, L145

\bibitem[Belloni \& Motta(2016)]{Belloni2016} Belloni, T.~M. \& Motta, S.~E.\ 2016, Astrophysics of Black Holes: From Fundamental Aspects to Latest Developments, 440, 61

\bibitem[Bondi \& Hoyle(1944)]{Bondi1944} Bondi, H. \& Hoyle, F.\ 1944, \mnras, 104, 273

\bibitem[Chernyakova et al.(2005)]{Chernyakova2005} Chernyakova, M., Courvoisier, T.~J.-L., Rodriguez, J., et al.\ 2005, The Astronomer's Telegram, 519

\bibitem[Cordova et al.(1981)]{Cordova1981} Cordova, F.~A., Mason, K.~O., \& Nelson, J.~E.\ 1981, \apj, 245, 609. 

\bibitem[Corradi et al.(1999)]{Corradi1999} Corradi, R.~L.~M., Ferrer, O.~E., Schwarz, H.~E., et al.\ 1999, \aap, 348, 978

\bibitem[de Val-Borro et al.(2017)]{deValBorro2017} de Val-Borro, M., Karovska, M., Sasselov, D.~D., et al.\ 2017, \mnras, 468, 3408

\bibitem[Dyson \& Williams(1997)]{Dyson1997} Dyson, J.~E. \& Williams, D.~A.\ 1997, The physics of the interstellar medium.  Edition: 2nd ed. Publisher: Bristol: Institute of Physics Publishing, 1997. Edited by J. E. Dyson and D. A. Williams. Series: The graduate series in astronomy. ISBN: 0750303069

\bibitem[Eze(2014)]{Eze2014} Eze, R.~N.~C.\ 2014, \mnras, 437, 857

\bibitem[Fekel et al.(2000)]{Fekel2000} Fekel, F.~C., Joyce, R.~R., Hinkle, K.~H., et al.\ 2000, \aj, 119, 1375. %doi:10.1086/301260

\bibitem[Gabriel et al.(2004)]{Gabriel2004} Gabriel, C., Denby, M., Fyfe, D.~J., et al.\ 2004, Astronomical Data Analysis Software and Systems (ADASS) XIII, 314, 759

\bibitem[Hric et al.(1998)]{Hric1998} Hric, L., Petrik, K., Urban, Z., et al.\ 1998, \aap, 339, 449

\bibitem[Ishida et al.(2009)]{Ishida2009} Ishida, M., Okada, S., Hayashi, T., et al.\ 2009, \pasj, 61, S77. %doi:10.1093/pasj/61.sp1.S77

\bibitem[Kellogg et al.(2007)]{Kellogg2007} Kellogg, E., Anderson, C., Korreck, K., et al.\ 2007, \apj, 664, 1079

\bibitem[Kellogg et al.(2001)]{Kellogg2001} Kellogg, E., Pedelty, J.~A., \& Lyon, R.~G.\ 2001, \apjl, 563, L151

\bibitem[Kennea et al.(2009)]{Kennea2009} Kennea, J.~A., Mukai, K., Sokoloski, J.~L., et al.\ 2009, \apj, 701, 1992
%. doi:10.1088/0004-637X/701/2/1992

\bibitem[Kenyon \& Garcia(1986)]{Kenyon1986} Kenyon, S.~J. \& Garcia, M.~R.\ 1986, \aj, 91, 125. %doi:10.1086/113991

\bibitem[Lee et al.(2022)]{Lee2022} Lee, Y.-M., Kim, H., \& Lee, H.-W.\ 2022, \apj, 931, 142

\bibitem[Linford et al.(2019)]{Linford2019} Linford, J.~D., Chomiuk, L., Sokoloski, J.~L., et al.\ 2019, \apj, 884, 8. %doi:10.3847/1538-4357/ab3c62

\bibitem[Liu et al.(2017)]{Liu2017} Liu, Z.-W., Stancliffe, R.~J., Abate, C., et al.\ 2017, \apj, 846, 117

\bibitem[Lodders(2003)]{Lodders2003} Lodders, K.\ 2003, \apj, 591, 1220

\bibitem[Lucy et al.(2020)]{Lucy2020} Lucy, A.~B., Sokoloski, J.~L., Munari, U., et al.\ 2020, \mnras, 492, 3107. 

%\bibitem[Luna(2019)]{Luna2019} Luna, G.~J.~M.\ 2019, Boletin de la Asociacion Argentina de Astronomia La Plata Argentina, 61, 93

\bibitem[Luna et al.(2018)]{Luna2018} Luna, G.~J.~M., Mukai, K., Sokoloski, J.~L., et al.\ 2018, \aap, 619, A61. %doi:10.1051/0004-6361/201833747

\bibitem[Luna et al.(2013)]{Luna2013} Luna, G.~J.~M., Sokoloski, J.~L., Mukai, K., et al.\ 2013, \aap, 559, A6 %doi:10.1051/0004-6361/201220792

\bibitem[Luna et al.(2008)]{Luna2008} Luna, G.~J.~M., Sokoloski, J.~L., \& Mukai, K.\ 2008, RS Ophiuchi (2006) and the Recurrent Nova Phenomenon, 401, 342

\bibitem[Merc et al.(2024)]{Merc2024} Merc, J., Beck, P.~G., Mathur, S., et al.\ 2024, \aap, 683, A84

\bibitem[Merc et al.(2019)]{Merc2019} Merc, J., G{\'a}lis, R., \& Wolf, M.\ 2019, Astronomische Nachrichten, 340, 598
%. doi:10.1002/asna.201913662

\bibitem[Motta et al.(2021)]{Motta2021} Motta, S.~E., Rodriguez, J., Jourdain, E., et al.\ 2021, \nar, 93, 101618

\bibitem[Mukai et al.(2007)]{Mukai2007} Mukai, K., Ishida, M., Kilbourne, C., et al.\ 2007, \pasj, 59, 177. % doi:10.1093/pasj/59.sp1.S177

\bibitem[Mukai(2017)]{Mukai2017} Mukai, K.\ 2017, \pasp, 129, 062001
%. doi:10.1088/1538-3873/aa6736

\bibitem[M\"{u}rset et al.(1997)]{Murset1997} Muerset, U., Wolff, B., \& Jordan, S.\ 1997, \aap, 319, 201

\bibitem[M{\"u}rset \& Schmid(1999)]{Murset1999} M{\"u}rset, U. \& Schmid, H.~M.\ 1999, \aaps, 137, 473. doi:10.1051/aas:1999105

\bibitem[Munari et al.(2016)]{Munari2016} Munari, U., Dallaporta, S., \& Cherini, G.\ 2016, \na, 47, 7

\bibitem[Mu{\~n}oz-Darias et al.(2014)]{MD2014} Mu{\~n}oz-Darias, T., Fender, R.~P., Motta, S.~E., et al.\ 2014, \mnras, 443, 3270

\bibitem[Nasa High Energy Astrophysics Science Archive Research Center (Heasarc)(2014)]{Heasoft2014} Nasa High Energy Astrophysics Science Archive Research Center (Heasarc)\ 2014, Astrophysics Source Code Library. ascl:1408.004

\bibitem[Orio et al.(2007)]{Orio2007} Orio, M., Zezas, A., Munari, U., et al.\ 2007, \apj, 661, 1105

\bibitem[Paltani \& Ricci(2017)]{Paltani2017} Paltani, S. \& Ricci, C.\ 2017, \aap, 607, A31. %doi:10.1051/0004-6361/201629623

\bibitem[Patterson \& Raymond(1985)]{Patterson1985} Patterson, J. \& Raymond, J.~C.\ 1985, \apj, 292, 535

\bibitem[Podsiadlowski \& Mohamed(2007)]{Podsiadlowski2007} Podsiadlowski, P. \& Mohamed, S.\ 2007, Baltic Astronomy, 16, 26

\bibitem[Pringle \& Savonije(1979)]{Pringle1979} Pringle, J.~E. \& Savonije, G.~J.\ 1979, \mnras, 187, 777

\bibitem[Pujol et al.(2023)]{Pujol2023} Pujol, A., Luna, G.~J.~M., Mukai, K., et al.\ 2023, \aap, 670, A32

\bibitem[Sacchi et al.(2024)]{Sacchi2024} Sacchi, A., Karovska, M., Raymond, J., et al.\ 2024, \apj, 961, 12

\bibitem[Schaefer(2010)]{Schaefer2010} Schaefer, B.~E.\ 2010, \apjs, 187, 275. doi:10.1088/0067-0049/187/2/275

\bibitem[Schaefer(2023)]{Schaefer2023}
Schaefer, B. E.\ 2023, Journal for the History of Astronomy, 54(4), 436-455

\bibitem[Shara et al.(2018)]{Shara2018} Shara, M.~M., Prialnik, D., Hillman, Y., et al.\ 2018, \apj, 860, 110

\bibitem[Sion et al.(2019)]{Sion2019} Sion, E.~M., Godon, P., Mikolajewska, J., et al.\ 2019, \apj, 874, 178

\bibitem[Soker(2016)]{Soker2016} Soker, N.\ 2016, \nar, 75, 1

\bibitem[Svoboda et al.(2017)]{Svoboda2017} Svoboda, J., Guainazzi, M., \& Merloni, A.\ 2017, \aap, 603, A127

\bibitem[Toal{\'a}, Botello \& Sabin(2023)]{Toala2023} Toal{\'a}, J.~A., Botello, M.~K., \& Sabin, L.\ 2023, \apj, 948, 14

\bibitem[Toal{\'a}(2024)]{Toala2024} Toal{\'a}, J.~A.\ 2024, \mnras, 528, 987

\bibitem[Toal{\'a} et al.(2023)]{Toala2023b} Toal{\'a}, J.~A., Gonz{\'a}lez-Mart{\'\i}n, O., Karovska, M., et al.\ 2023, \mnras, 522, 6102

\bibitem[Toal{\'a} et al.(2022)]{Toala2022} Toal{\'a}, J.~A., Sabin, L., Guerrero, M.~A., et al.\ 2022, \apjl, 927, L20

\bibitem[Tueller et al.(2005)]{Tueller2005} Tueller, J., Barthelmy, S., Burrows, D., et al.\ 2005, The Astronomer's Telegram, 668

\bibitem[Webb(2023)]{Webb2023} Webb, N.~A.\ 2023, arXiv:2303.10055. doi:10.48550/arXiv.2303.10055

\bibitem[Yu et al.(2022)]{Yu2022} Yu, Z.-. li ., Xu, X.-. jie ., Shao, Y., et al.\ 2022, \apj, 932, 132

\bibitem[Zamanov et al.(2024)]{Zamanov2024} Zamanov, R.~K., Stoyanov, K.~A., Marchev, V., et al.\ 2024, arXiv:2405.11506. doi:10.48550/arXiv.2405.11506

\bibitem[Zamanov et al.(2023)]{Zamanov2023} Zamanov, R., Boeva, S., Latev, G.~Y., et al.\ 2023, \aap, 680, L18

\bibitem[Zhekov \& Tomov(2019)]{Zhekov2019} Zhekov, S.~A. \& Tomov, T.~V.\ 2019, \mnras, 489, 2930. %doi:10.1093/mnras/stz2329

\end{thebibliography}

\section*{DATA AVAILABILITY}

The X-ray and optical data underlying this article were retrieved from publicly available archives. The processed data will be shared on reasonable request to the corresponding author.

\appendix

\section{Temporal evolution in optical bands}
\label{sec:app}

\begin{figure*}
\begin{center}
\includegraphics[angle=0,width=0.95\linewidth]{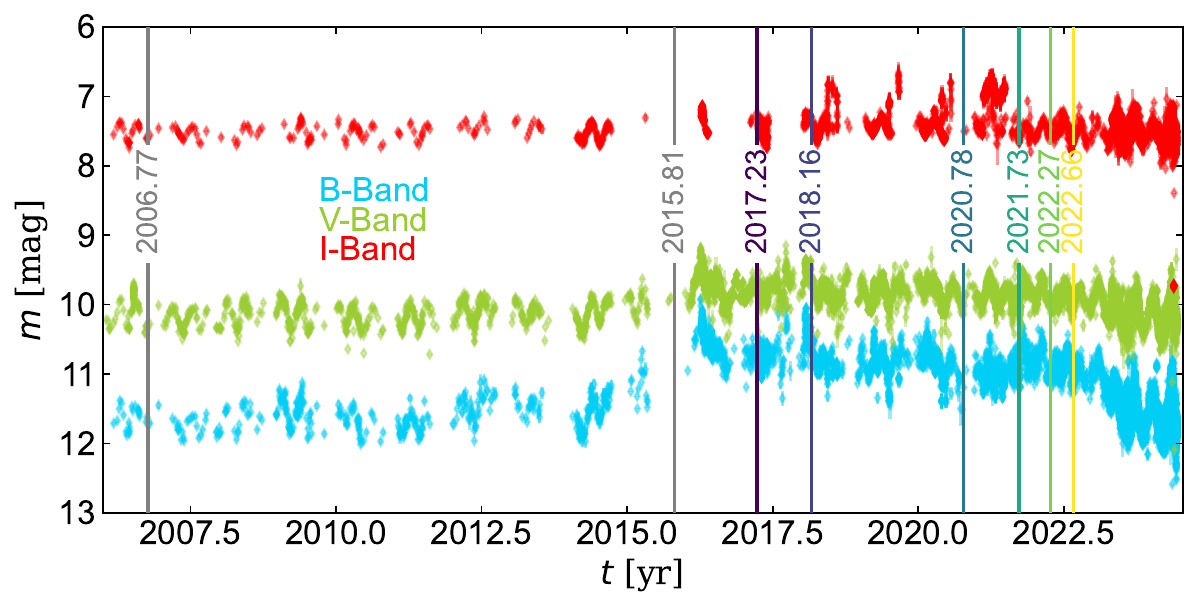}
\caption{Light curve of T~CrB obtained from photometric measurements obtained with the $I$, $V$ and $B$ filters from the AAVSO. The vertical lines represent the epochs in which {\it XMM-Newton} observations are available and analysed in this paper.}
\label{fig:VBI}
\end{center}
\end{figure*}

Fig.~\ref{fig:VBI} presents the temporal evolution of the optical light curves of T~CrB obtained thought the $I$, $V$ and $B$ filters obtained from AAVSO.

% Don't change these lines
\bsp	% typesetting comment
\label{lastpage}

\end{document}